\documentclass[a4paper,11pt]{article}
\pdfoutput=1 

\usepackage{jcappub} 
                     
\usepackage{aas_macros} 

\usepackage[T1]{fontenc} 
\usepackage[utf8]{inputenc}

\usepackage[english]{babel}    
\usepackage{xspace}
\usepackage{microtype}    
\usepackage[utf8]{inputenc}    

\usepackage[T1]{fontenc}
\usepackage[extramarks]{titleps}
\usepackage{lipsum}

\usepackage{amsmath,amssymb,amsfonts,amscd,dsfont,bm}     
\usepackage{amsthm}
\allowdisplaybreaks     
\usepackage{mathrsfs}
\usepackage[usenames,dvipsnames]{xcolor}
\usepackage{slashed}
\usepackage{youngtab}
\usepackage{physics}
\usepackage{orcidlink}
\usepackage[capitalise]{cleveref}
\allowdisplaybreaks

\usepackage{natbib}
\usepackage{hyperref}

\usepackage{graphicx}
\graphicspath{{./Figures/}}

\crefformat{equation}{Eq.~(#2#1#3)}
\crefformat{table}{Tab.~(#2#1#3)}
\crefformat{figure}{Fig.~(#2#1#3)}
\crefformat{appendix}{App.~(#2#1#3)}
\crefformat{section}{Sec.~(#2#1#3)}
\crefformat{subsection}{Subsec.~(#2#1#3)}

\usepackage[normalem]{ulem}

\title{
\huge
Global stability of ghostly field theories:\\
Classical scattering in $(N+1)$ dimensions
}

\author[a]{Aaron Held\,\orcidlink{0000-0003-2701-9361}}
\affiliation[a]{
Institut de Physique Théorique Philippe Meyer, Laboratoire de Physique de l’\'Ecole normale sup\'erieure (ENS), Universit\'e PSL, CNRS, Sorbonne Universit\'e, Universit\'e Paris Cité, F-75005 Paris, France
}
\emailAdd{aaron.held@phys.ens.fr}

\abstract{
We review results on small-data global stability and highlight their applicability to classical interacting scalar fields with opposite-sign kinetic terms (ghosts).
Further, we present a one-parameter family of numerical scattering solutions under the simplifying assumption of spherical symmetry. We thereby identify classical ghostly field theories with polynomial interaction potentials for which small-data global stability apparently extends to large-data global stability, i.e., global stability for \emph{all} compactly supported initial data. 
These global stability results support an underlying physical mechanism by which the ghost instability can be quenched due to dominant self-interactions, at least in classical physics.
}
	
\begin{document}
\setcounter{tocdepth}{2}
\maketitle


\section{Introduction}
\label{sec:introduction}

\subsection*{Motivation: Are all higher-derivative systems catastrophically unstable?}

The Ostrogradski theorem~\cite{Ostrogradsky:1850fid} is often regarded as one of the most powerful restrictions on fundamental physics: It is frequently invoked as a stringent no-go statement to dismiss higher-order time derivatives as unphysical or, at least, as non-fundamental.
At the same time, higher derivatives appear naturally in effective field theory (EFT) and are induced by generic quantum corrections. Their dismissal thus has wide-ranging implications for modifications of gravity (see e.g.,~\cite{Lovelock:1971yv}), for cosmology (see e.g.,~\cite{Cline:2003gs,Carroll:2003st,Kaplan:2005rr}), and for quantum gravity (see e.g.,~\cite{Stelle:1976gc}).
Vice versa, the absence of higher-order degrees of freedom is widely applied as a theoretical prior in the search for new physics, see, e.g.,~\cite{Copeland:2006wr,Sotiriou:2008rp,Clifton:2011jh,deRham:2014zqa,Berti:2015itd} for reviews. 

The original formalism by Ostrogradski~\cite{Ostrogradsky:1850fid} assumes a non-degenerate point-particle Lagrangian with second-order time derivatives and constructs a suitable Hamiltonian, now known as the Ostrogradski Hamiltonian. This construction shows that the total conserved energy is unbounded in phase space, both from above and below. In fact, upon diagonalisation of the kinetic part of the Hamiltonian, one identifies opposite-sign kinetic terms -- the so-called Ostrogradski ghosts. Ostrogradski's formalism generalises to higher-order time derivatives and to classical field theory, see, e.g., ~\cite{deUrries:1998obu}. While the unboundedness of energy is a mathematical fact, the inevitability of a catastrophic instability is not. Instead, the latter is an expression of widespread physical expectations about opposite-sign kinetic terms, see~\cite{Woodard:2015zca} for a review. 
Several recent results -- briefly reviewed in the next paragraph -- stand in tension with these expectations and, hence, challenge the catastrophic nature and/or inevitability of an instability, at least in classical physics. In light of the extraordinary significance to fundamental physics (see above), we therefore consider it an extremely worthwhile scientific endeavour to carefully reassess the classical dynamics associated with opposite-sign kinetic terms. 

\subsection*{Context: Brief summary of previous work}

The stability of point particles with opposite-sign kinetic terms has a longstanding history (see, e.g.,~\cite{Pais:1950za,Lee:1970iw,Bender:2007wu}). 
It is useful to distinguish different notions of stability, in particular, with regard to whether stability holds in a local region of phase space (Lyapunov stability) or for the global set of all possible initial conditions (Lagrange stability). All such notions are typically global in time, i.e., required to hold for all future time. Since the purpose of the present paper is to address the stability of classical field theories, we review the literature on point particles with said purpose in mind, see~\cite{ErrastiDiez:2024hfq} for a more encompassing account. 

It seems widely accepted that, for point-particle systems with a finite-dimensional phase space, opposite-sign kinetic terms are not necessarily an obstruction to Lyapunov stability, see~\cite{Pagani:1987ue} for an early account. Numerical studies~\cite{Pavsic:2016ykq} support the related notion of ``islands of stability''~\cite{Smilga:2004cy}, i.e., finite regions of phase space in which trajectories remain confined for all future time.

In contrast, Lagrange stability implies that the evolution of all initial conditions remains bounded for all future time. Proof of Lagrange stability has -- for now -- only been possible in integrable systems, where additional constants of motion allow us to analytically solve the initial value problem, or at least to rephrase it in terms of closed-form integrals. First results in this direction rely on the presence of a bi-Hamiltonian structure: Bi-Hamiltonian systems are always integrable; in addition, they exhibit two distinct Poisson structures, hence, two distinct Hamiltonians, which generate the exact same equations of motion~\cite{Magri:1977gn}. In this setting, one bounded Hamiltonian can be used to establish Lagrange stability of another unbounded (ghostly) Hamiltonian, see e.g.~\cite{Kaparulin:2014vpa,Abakumova:2018eck}. It has been argued~\cite{ErrastiDiez:2025mgu} that such systems -- since they exhibit one bounded Hamiltonian -- should not be regarded as ``ghost-ridden'' systems in the first place. Certainly, their stability appears to require substantial additional structure. 

In any case, an independent proof strategy~\cite{Deffayet:2021nnt} does not rely on such bi-Hamiltonian structure\footnote{We expect that an application of Theorem 3.1 in~\cite{fernandes1994completely} can provide proof that the systems considered in~\cite{Deffayet:2021nnt,Deffayet:2023wdg} are \emph{not} of bi-Hamiltonian nature, at least not in the sense of Magri~\cite{Magri:1977gn}.}. Integrability alone has been sufficient to prove Lagrange stability in a systematic class of ghostly systems~\cite{Deffayet:2023wdg}, including a tower of polynomial interactions. 

Even when global stability cannot be established, i.e., an unquenched ghost instability leads to runaway solutions, a distinction between \emph{benign} and \emph{catastrophic} runaways has been emphasised:
An instability is referred to as ``catastrophic'' if it can lead to finite-time divergences in physical quantities. By contrast, if all runaway solutions can be extended to arbitrarily large time, the instability is referred to as ``benign''. This distinction has been pointed out, both for point-particle systems~\cite{Smilga:2004cy,Pavsic:2016ykq} and for classical field theory~\cite{Damour:2021fva,Deffayet:2025lnj}.

For the purpose of generalising to field theory, the recent analysis~\cite{Deffayet:2023wdg} of point-particle systems for two position-phase-space variables $x$ and $y$, with diagonal and opposite-sign kinetic terms and an interaction potential $V(x,y)$, is most relevant. In particular, it provides an underlying physical explanation: Lagrange stability occurs whenever self-interaction potentials $V_P(x)$ and $V_G(y)$ -- each stable in relation to its respective kinetic term -- are ``sufficiently dominant'' in comparison to the coupled interactions $V_\text{int}(x,y) := V(x,y) - V_P(x) + V_G(y)$, at least outside of a finite region of phase space. 
The term ``sufficiently dominant'' has been made precise in the case of polynomial potentials where it suffices\footnote{A somewhat weaker condition on the Hessian is also explored in~\cite{Deffayet:2023wdg}.} to demand that $V_\text{int}(x,y)$ is of lower polynomial degree than, both, $V_P(x)$ and $V_G(y)$.
Remarkably, a numerical study of random initial conditions suggests that the same holds true in absence of integrability~\cite{Deffayet:2023wdg}. Part of the motivation for the present work is to determine whether these physical insights into point-particle systems can be generalised to classical field theory.
\\

While nonlinear wave equations are a rich topic of mathematical analysis (see, e.g.,~\cite{sogge1995lectures,shatah2000geometric,tao2006nonlinear} or the dedicated review in~\cref{sec:analytical}), classical field theories with opposite-sign kinetic terms have for a long time been left mostly unattended\footnote{We can only explain this historical blind spot as a result of widespread negative expectations.} (see~\cite{Gross:2020tph,Damour:2021fva} for notable exceptions). More recently, however, they have caught the interest of numerical relativists due to their significance for well-posed nonlinear time evolution in the presence of higher-curvature corrections~\cite{Noakes:1983,Held:2021pht,Held:2023aap,Figueras:2024bba,Held:2025ckb}.

Motivated also by this practical use case, ghostly field theories have recently been investigated in (1+1) dimensions, focusing on periodic boundary conditions and non-compact plane-wave initial data~\cite{Deffayet:2025lnj}. Remarkably, the numerical evolution of field-theoretic generalisations of the above globally stable point-particle systems seem to be longlived: If any instability remains, it seems to be driven by repeated interactions. Moreover,~\cite{Deffayet:2025lnj} clarifies a decoupling mechanism previously observed in the context of quadratic gravity~\cite{Held:2023aap}: Contrary to prior expectations~\cite{Woodard:2015zca}, high-frequency modes are found to be more stable than low-frequency modes and increasingly heavy mass terms seem to quench the instability. Hence, as is explicitly demonstrated by an example~\cite{Deffayet:2025lnj}, massive high-frequency modes can be integrated out consistently. Shortly thereafter, the same decoupling has also been demonstrated for an EFT expansion of a ghost-free ultraviolet theory~\cite{Figueras:2025gal}: On the basis of small-data global existence theorems (see, e.g.,~\cite{sogge1995lectures}), the authors of~\cite{Figueras:2025gal} also formulate a conjecture for nonlinear stability of the EFT vacuum.

The present work can be understood as a continuation -- or rather a companion paper -- of~\cite{Deffayet:2025lnj}. In particular, it applies and extends the same numerical methods.
However, while~\cite{Deffayet:2025lnj} investigates $(1+1)$ dimensions, periodic boundary conditions, and mostly focuses on non-compact initial data~\cite{Deffayet:2025lnj}, the present work generalises to $(N+1)$ dimensions subject to spherical symmetry, open boundary conditions, and focuses on compact initial data to address scattering theory. As we will see, both the higher number of spatial dimensions as well as the focus on compact initial data and open boundary conditions, work in favour of stability.

In the context of EFT and in relation to the conjecture in~\cite{Figueras:2025gal}, the present work clarifies that existing small-data global existence theorems (see~\cite{klainerman1985,simon1993} and subsequent review in~\cref{sec:math:small-data}) already establish nonlinear vacuum stability for a subclass of non-derivative models as detailed below. We anticipate that this will be helpful when considering extensions that include derivative interactions.

\subsection*{Setup: Scalar field theories with non-deriative interactions}

Throughout this work, we consider $(N+1)$ dimensional classical field theories. Given our motivation to generalise the understanding of global stability developed in~\cite{Deffayet:2023wdg} from point-particle systems to field theory, we restrict to two scalar fields $\phi(t,\vec{x})$ and $\chi(t,\vec{x})$, interacting through a non-derivative potential $V[\phi,\chi]$.
Depending on a parameter $\sigma=\pm1$, the two scalar fields can be chosen to have opposite-sign kinetic terms. To be specific, we consider the associated Lagrangian density\footnote{We use the mostly-plus signature convention.}
\begin{align}
\label{eq:field-theory-Lagrangian}
	\mathcal{L} = 
	- \frac{1}{2}\phi\left[\Box - m_\phi^2\right]\phi
	-  \frac{\sigma}{2}\chi\left[\Box - m_\chi^2\right]\chi
	+ V[\phi,\,\chi]\;.
\end{align}
Further, we mirror the decomposition of the potential into three parts, following the point-particle analysis in~\cite{Deffayet:2023wdg}, see also~\cite{Deffayet:2025lnj}, i.e.,
\begin{align}
    V_\phi[\phi] &= V[\phi,\,\chi=0]\;,
    \notag\\
    V_\chi[\chi] &= \sigma\,V[\phi=0,\,\chi]\;,
    \notag\\
    V_\text{int}[\phi,\,\chi] &= V[\phi,\,\chi] - V_\phi[\phi] - \sigma\,V_\chi[\chi]\;.
\end{align}
We refer to $V_\phi$ and $V_\chi$ as self-interaction potentials and to $V_\text{int}$ as the ghostly interaction potential.
For concreteness, we work with polynomial interactions of the form
\begin{align}
    V_\phi[\phi] &= \lambda_{\ell}\,\phi^\ell\;,
    \label{eq:Vphi}
    \\
    V_\chi[\chi] &= \lambda_{\ell}\,\chi^\ell\;,
    \label{eq:Vchi}
    \\
    V_\text{int}[\phi,\,\chi] &= \lambda_{mn}\,\phi^m\,\chi^n\;,
    \label{eq:Vcross}
\end{align}
where we will restrict to positive integer exponents $\ell,m,n\in\mathbb{N}$.
The respective 2nd-order field equations can be written as
\begin{alignat}{9}
	\left[\Box - m_\phi^2\right]\phi 
    &=&
    &+ \partial_{\phi} V 
    &=
    &+ \partial_{\phi} V_\phi&
    &+ \partial_{\phi} V_\text{int}\;
    &\;\equiv\;& 
    F_\phi(\phi,\chi)
    \;,
	\notag\\
	\left[\Box - m_\chi^2\right]\chi 
    &=&
    &+ \sigma \,\partial_{\chi} V
    &=
    &+ \partial_{\chi} V_\chi&
    &+ \sigma \,\partial_{\chi} V_\text{int}
    &\;\equiv\;&
    F_\chi(\phi,\chi)
    \;.
	\label{eq:eoms}
\end{alignat}
The notation $F_\phi(\phi,\chi)$, $F_\chi(\phi,\chi)$ indicates the nonlinearities of the system at the level of evolution equations to conform with typical mathematical notation. 

Written in this form, it is suggestive that classic results in mathematical analysis of nonlinear hyperbolic differential equations apply, irrespective of the sign of $\sigma=\pm1$, hence, irrespective of the presence of a ghost. In particular, the above evolution equations constitute a system of two coupled nonlinear wave (for $m_\phi= 0$, $m_\chi= 0$) or Klein-Gordon (for $m_\phi\neq 0$, $m_\chi\neq 0$) equations with non-derivative nonlinearities $F_\phi(\phi,\chi)$, $F_\chi(\phi,\chi)$. The choice of $\sigma=\pm1$ concerns only the sign of these nonlinear terms.
Since the latter do not contain derivatives, $\sigma=\pm1$ does not affect the local solution (local well-posedness). Moreover, as will be the main topic of this work, even global solutions (global well-posedness) can apparently exist.

\subsection*{Summary: Outline of the remaining paper}

The remainder of this paper makes three important points:
\begin{itemize}
    \item 
    In~\cref{sec:analytical}, we review the mathematical literature on global stability of wave-type equations, clarifying, in particular, that established small-data global stability theorems~\cite{klainerman1985,simon1993,shatah1985normal,simon1985,klainerman1986null,georgiev1990global} apply to~\cref{eq:eoms}, provided that the nonlinearities are of sufficiently high degree when expanded around the respective vacuum. The specific admissible degree depends on the presence of mass-terms and on the spatial dimension $N$ (see~\cref{sec:analytical}): For instance in $(3+1)$ dimensions and with non-vanishing and distinct mass terms, it suffices that $F_\phi(\phi,\chi)$ and $F_\chi(\phi,\chi)$ start at quadratic order~\cite{klainerman1985,simon1993}.
    We highlight that small-data global stability concerns all compactly supported initial data with sufficiently small amplitude. Applicability of these theorems therefore directly implies nonlinear stability of the vacuum against all sufficiently small and localised perturbations -- even in the presence of ghosts. 
    \item 
    In~\cref{sec:numerical}, we turn to the numerical scattering solutions in spherical symmetry. The analysis suggests that small-data global stability apparently extends to large-data global stability, i.e., to global stability for \emph{all} compactly supported initial data, provided that, at large data, stable self-interactions dominate over ghostly interactions.
    \item 
    The above insight essentially mirrors insights for point-particle systems~\cite{Deffayet:2023wdg}. Remarkably, while we see no obvious route to a proof of Lagrange stability for point particle systems in absence of integrability, the tools of mathematical analysis seem to provide a feasible route to do so in classical field theory. From a physics perspective, this seems to be routed in dissipation (or geometric spreading) which stabilises the classical field theory in comparison to point particle systems. In fact, the same holds for a single positive-energy degree of freedom. As has been discussed also in~\cite{Figueras:2025gal}, point particle motion in an unstable potential (e.g., $V(x)=-x^4$) is unstable for all non-trivial initial conditions, while the respective field theory (e.g., with $V(\phi)=-\phi^4$ in (3+1) dimensions) nevertheless enjoys small-data global stability. 
    The same physical mechanism also stabilises ghostly field theories.
    In~\cref{sec:expectations}, we formulate resulting expectations for physical stability. 
\end{itemize}
We conclude in~\cref{sec:discussion} by discussing potential extensions and implications, in particular, with regards to derivative interactions and quantum stability.

\newpage

\section{Applicability of small-data global stability theorems}
\label{sec:analytical}

The purpose of this section is twofold. 
On the one hand, we provide a concise physicists review of the key aspects of global stability for nonlinear hyperbolic partial differential equations (PDEs), focusing in particular on physical insights behind the proof strategy. We review, both, small-data global stability and large-data global stability. On the other hand, we use the opportunity to emphasise that the proven small-data global stability theorems apply directly to the ghostly field theories at hand.

\subsection{Physical intuition behind the mathematical proof strategies}
\label{sec:physics-review-of-PDE-theory}

We find it helpful to review the respective physical intuition behind the mathematical proofs, hoping to provide a useful dictionary between the languages of mathematicians and physicists.

Physicists tend to distinguish the sign of non-derivative interaction terms at the level of the potential and refer to potentials as stable (or unstable) if they are bounded (or unbounded) from below with respect to the corresponding kinetic energy\footnote{For a positive kinetic-energy field, a stable potential is bounded from below, but for a negative kinetic-energy field, a stable potential needs to be bounded from above.}.
Mathematicians distinguish the sign of nonlinearities at the level of the field equations and refer to them as focusing or defocusing, depending on whether they tend to concentrate or disperse the kinetic energy of a localised field configuration. At least for nonlinear wave equation with non-derivative interactions that stem from an action, focusing nonlinearities correspond to unstable potentials and defocusing nonlinearities correspond to stable potentials.

\subsubsection{On the role of positive-definite potential energy}

For our purposes, it is useful to distinguish global stability theorems depending on whether or not their proofs require to make a positive-definite energy assumption. When the potential energy is positive definite (in comparison to the respective kinetic energy), local energy inequalities can be established. While the field theory can still convert potential to kinetic energy and vice versa, the total energy budget is conserved. Assuming positive-definite potential energy (coercivity assumption in the mathematical literature) thus guarantees that these energy inequalities are valid if the field values tend to infinity. 

Coercivity assumptions and resulting energy inequalities have been used to establish global existence results, see~\cref{sec:math:large-data} as well as, e.g.,~\cite{sogge1995lectures,shatah2000geometric} for lectures summarising this literature. Our field equations (for the ghostly $\sigma=-1$ case) do not allow us to make such a coercivity assumption and so we cannot directly make use of this class of theorems.

However, as also highlighted in~\cite{Figueras:2025gal}, theorems on small-data global existence do not typically require a coercivity assumption: Local energy estimates are still necessary but can instead be achieved relying solely on the small-data assumption and the resulting smallness of the nonlinear contributions in comparison to the geometric (and dispersive in the presence of mass terms) spreading into the available spatial dimensions.

\subsubsection{Geometric spreading and decay estimates for small-data global stability}
\label{sec:geometric-dissipative-decay}
\begin{table*}[]
	\centering
    {\renewcommand{\arraystretch}{1.5}
    \begin{tabular}{c||c|c|c}
    		 & spatial dimension $N$ & critical exponent $p_\text{crit}$ & reference
    		\\\hline\hline
    		wave & $N=1$ & -- & 
    		\\\hline
    		wave & $N\geqslant 2$ & $\frac{N + 1 + \sqrt{(N + 1)^2 + 8(N - 1)}}{2(N - 1)}$
            & see~\cite{strauss1981nonlinear}
    		\\\hline\hline
    		Klein-Gordon & $N\geqslant 1$ & $2/N$ & see~\cite{segal1963global,ginibre1985global}
    \end{tabular}
    }
    \caption{
    \label{tab:critical-exponents} 
    Summary of small-data critical exponents $p_\text{crit}$ for the massless semilinear wave equation $\Box\phi = \pm |\phi|^p$ and Klein-Gordon equation $(\Box-m_\phi^2)\phi = \pm |\phi|^p$ in $(N+1)$ dimensions, including references to the original mathematical literature. For $p>p_\text{crit}(N)$, the listed mathematical references establish \emph{small-data} global stability for compactly supported initial data.
    }
\end{table*}

It is crucial that the linear (free) wave equation leads to geometric spreading: A localised initial field profile tends to spread out into the available spatial dimensions. As intuition suggests: The higher the number of dimensions, the faster this decay. In the presence of (non-tachyonic) mass terms, there is an additional effect of dispersion which also leads to faster decay. The decay for the linear wave equation and the linear Klein-Gordon equation, respectively, thus depends on the number of spatial dimensions $N$ and is given by
\begin{alignat}{4}
	\Box\phi &= 0\;
	\quad &\Rightarrow\quad
	&|\phi(t,\vec{x})|\sim t^{-(N-1)/2}
	\quad &\text{for large t}
	\;,
	\\
	(\Box - m_\phi^2)\phi &= 0\;
	\quad &\Rightarrow\quad
	&|\phi(t,\vec{x})|\sim t^{-N/2}
	\quad &\text{for large t}
	\;.
\end{alignat}
We note that a higher number of spatial dimensions leads to faster decay (due to geometric spreading) and that the presence of a mass term leads to faster decay (due to dispersion). Physically, the Klein-Gordon decay can be understood simply because the localised solution will disperse into a volume of size $\sim t^N$ at large times $t$, where ``large'' refers to a comparison with the initial localised region. Without a mass term, the wave does not disperse but rather propagates on the light cone. It therefore spreads into a spherical shell, i.e., into a surface volume~$\sim t^{N-1}$, leading to slower decay. 

\subsection{Review of global stability results for small data}
\label{sec:math:small-data}

For small data (controlled by some sufficiently small $\epsilon$), it is possible to show that the above geometric/dispersive decay dominates any nonlinearity $\pm|\phi|^p$ of sufficiently high degree $p$. Sharp results for the single-field wave and Klein-Gordon equations have been established and are reviewed below. The result is summarised in~\cref{tab:critical-exponents}.

The required decay estimates rely on the Duhamel formula (see, e.g.,~\cite{sogge1995lectures, tao2006nonlinear}), which decomposes the full solution into a free part $\phi_{\mathrm{free}}$ and an inhomogeneous part, depending on the nonlinearity $F(\phi)$. This is directly analogous to decomposing the solution of an ordinary differential equation (ODE) into homogeneous and inhomogeneous parts. In terms of the retarded Green's function $G_{\mathrm{ret}}$ of the free wave or Klein--Gordon equation in $(N+1)$ dimensions, the Duhamel formula reads
\begin{align}
  \phi(T,\vec{x}) = \phi_{\mathrm{free}}(T,\vec{x})
  + \int_0^T \int_{\mathbb{R}^N} G_{\mathrm{ret}}(T-t, \vec{x}-\vec{y})\,
      F(\phi(t,\vec{y})) \; d^Ny\; dt .
      \label{eq:Duhamel}
\end{align} 
Formally,~\cref{eq:Duhamel} holds for any nonlinearity $F(\phi)$: The analytical challenge is whether the spacetime integral converges for large times, relating precisely to the distinction between global existence and finite-time divergences.
From~\cref{eq:Duhamel}, the critical exponent $p$ can thus heuristically be determined by demanding that the time integral converges. 
\\

For the massive (Klein-Gordon) case and a nonlinearity of degree $p$, the integral of the nonlinearity can be controlled by the linear decay estimate, i.e.,
\begin{align}
	\int_0^T |\phi|^p
	<
	\int_0^T\,t^{-p\, N/2}\,dt
	\;,
\end{align}
which converges for large $T\rightarrow\infty$ if $p>p_\text{crit}$ with critical exponent
\begin{align}
	p_\text{crit}^\text{KG}(N) = \frac{2}{N}
	\;.
\end{align}
This critical exponent $p_\mathrm{crit}^\mathrm{KG}$ has been rigorously proven and shown to be sharp: small-data global existence holds for $p>p_\mathrm{crit}^\mathrm{KG}$~\cite{segal1963global,ginibre1985global}, 
while divergent solutions exist for $p \leqslant p_\mathrm{crit}^\mathrm{KG}$~\cite{glassey1981finite}.
\\

For the massless wave equation, the situation is somewhat more subtle because the linear decay of the free solution is slower, i.e.,  $|\phi|\sim t^{-(N-1)/2}$ in $N\geqslant 2$ without dissipation, instead of $|\phi|\sim t^{-N/2}$ with dissipation, see~\cref{sec:geometric-dissipative-decay}.
Pointwise decay estimates are insufficient to control $\phi$ itself. Rather, they can only control derivatives of the full solution. 
More refined methods (requiring mixed norms, see, e.g.,~\cite{sogge1995lectures, tao2006nonlinear}) lead to the critical Strauss exponent (conjectured in~\cite{strauss1981nonlinear})
\begin{align}
	p_\text{crit}^\text{wave}(N\geqslant 2) &= p_\text{Strauss} = 
    \frac{N + 1 + \sqrt{(N + 1)^2 + 8(N - 1)}}{2(N - 1)}
	\;,
    \label{eq:Strauss-exponent}
\end{align}
which is strictly larger than $p_\text{crit}^\text{KG}(N)$, confirming the physical intuition that the lack of dissipation makes it harder to establish global stabilty.
Rigorous results confirm that the Strauss exponent is sharp: for $p \leqslant p_\text{Strauss}$, finite-time divergences occur~\cite{john1981blow,glassey1981finite,sideris1984nonexistence,schaeffer1985equation}, while small-data global stability holds for $p > p_\text{Strauss}$~\cite{lindblad1995existence,lindblad1996long,georgiev1997weighted,keel1998endpoint}. 
In one spatial dimension $N=1$, the decay of the free solution is insufficient and small-data global stability for the massless wave equation cannot be established.
\\

The above small-data critical exponents are summarised in~\cref{tab:critical-exponents}. As reviewed, they are sharp for the single field case.
The same balance between nonlinear interactions and geometric/dissipative extends also to coupled systems. 
For coupled systems, the critical exponent of the single-field nonlinearity $p_\text{crit}$ is replaced by a sufficiently high degree $q$ in a polynomial expansion of the multi-field nonlinearity around the respective vacuum. 
We caution that $q$ cannot directly be identified with $p_\text{crit}$. Moreover, as one may expect from the distinction between geometric and dissipative decay, established results typically depend on the mass spectrum of the coupled system~\cite{klainerman1985,simon1993,shatah1985normal,simon1985,klainerman1986null,georgiev1990global}.

Nevertheless, for instance, in (3+1) dimensions small-data global stability is established~\cite{klainerman1985,simon1993} for compact initial data and arbitrary coupled systems of Klein-Gordon equations with non-vanishing and distinct masses if the nonlinearity starts at quadratic order or higher. (See~\cite[p.~434]{simon1993} for an explicit statement of applicability to the coupled case.)
\\

We close this section by highlighting once more that these type of decay estimates do not rely on any conserved quantity or Hamiltonian structure and hold, irrespective of the focusing/defocusing nature of the nonlinearity. 
This implies that small-data global stability results also hold irrespective of $\sigma=\pm1$, i.e., even in the presence of a ghost.

\subsection{Review of global stability results for large data}
\label{sec:math:large-data}

For a discussion of large-data global stability, we focus on a single massless semi-linear wave equation $\Box\phi = \pm |\phi|^{P-1}\phi$. In contrast to small-data global stability, large-data global stability proofs make use of the conserved total energy (Hamiltonian). 
As we will see, large-data global stability hinges on another critical exponent, referred to as the energy-critical exponent and here denoted by capital $P_\text{crit}(N)$ to destinguish from the small-data critical exponent $p_\text{crit}(N)$ of the previous section. The energy-critical exponent $P_\text{crit}(N)$ is determined as the special exponent for which the scaling symmetry of the evolution equation is also a scaling-symmetry for the total energy (Hamiltonian). It is independent of the presence/absence of mass terms.

Under a general scale transformation
\begin{align}
\phi(t,x) \mapsto \lambda^{\alpha} \phi(\lambda t, \lambda x).
\end{align}
the wave operator transforms like $\Box\phi\mapsto\lambda^{2+\alpha}\Box\phi$ while the nonlinearity transforms like $|\phi|^{P-1}\phi\mapsto \lambda^{P\alpha}|\phi|^{P-1}\phi$. Hence, the semi-linear wave equation is invariant under a scale transformation with
\begin{align}
    \alpha = \frac{2}{P-1}
    \;.
    \label{eq:scaling}
\end{align}
In contrast, the conserved total energy (Hamiltonian) is given by\begin{align}
H = 
\underbrace{\frac{1}{2} \int_{\mathbb{R}^N} \left( |\partial_t \phi|^2 + |\nabla \phi|^2 \right) \, dx}_\text{kinetic (and gradient) energy $T$}
\underbrace{\pm\frac{1}{p+1} \int_{\mathbb{R}^N} |\phi|^{P+1} \, dx}_\text{potential energy $V$}.
\end{align}
Here, the first integral denotes the kinetic energy (or, to be precise, the sum of kinetic and gradient energy) denoted by $T$. The second integral denotes the potential energy $V$ of the nonlinear interaction. Under the scale transformation, $T\mapsto \lambda^{2(\alpha+1)-N}T$ while $V\mapsto \lambda^{\alpha(P+1)-N}V$. However, scale invariance of the total energy requires not just that the exponents match (which reproduces~\cref{eq:scaling}) but rather that both exponents vanish, hence, that
\begin{align}    
    P_\text{crit}(N) = 1+\frac{4}{N-2}
    \;.
\end{align}
The case of $P<P_\text{crit}$ is referred to as energy subcritical; the case of $P=P_\text{crit}$ as energy critical; and the case of $P>P_\text{crit}$ as energy supercritical. The notion of energy criticality turns out to be crucial for large-data global stability.
In particular, for the defocusing and energy (sub-) critical case, it has been possible to prove large-data global stability for the semi-linear wave equation, i.e., for $\Box\phi = \pm |\phi|^{P-1}\phi$ and $P\leq P_\text{crit}(N)$.
\\

Rigorous results for large-data global well-posedness and scattering have been established in a collective effort: 
In the energy-subcritical regime ($P<P_\text{crit}(N)$), global existence follows from energy conservation and mixed-norm estimates~\cite{ginibre1995generalized,lindblad1995existence}. The energy-critical case ($P=P_\text{crit}(N)$) has been addressed in~\cite{grillakis1990regularity,grillakis1992regularity} and completed in~\cite{shatah1994well,shatah2000geometric}. 
Finally,~\cite{KenigMerle2006wave, kenig2008global} establish sharp thresholds with an exact criterion on the critical compactness of initial data. 
For comprehensive modern account, see also~\cite{tao2006nonlinear}.

\subsection{A comment on derivative nonlinearities}
\label{sec:small-data:derivatives}

The mathematical community has been able to establish much more general results for small-data global stability including derivative nonlinearities, see, for instance,~\cite{klainerman1985,simon1985,simon1993} for results concerning systems of Klein-Gordon equations with derivative nonlinearities. These results have been influential also for proofs of nonlinear stability (i.e., small-data global stability) of Minkowski spacetime~\cite{klainerman1986null, christodoulou1986global} and Schwarzschild spacetime~\cite{dafermos2021non} in General Relativity. A proof of nonlinear stability of Kerr spacetime remains outstanding, at least in the full subextremal range of spin, see, e.g.,~\cite{klainerman2022brief,dafermos2024quasilinear} and references therein. 

We note that, for instance, the results in~\cite{klainerman1985}, see also~\cite{simon1985,simon1993} for an explicit discussion of its applicability to coupled systems of Klein-Gordon equations, are quite general. They establish a small-data global stability statement for nonlinear Klein-Gordon systems
\begin{align}
    \Box\Phi + M_\Phi^2\,\Phi = F(\Phi, \Phi', \Phi'')
\end{align}
as long as the nonlinearities in $F$ only start at quadratic order or higher. Here, $\Phi$ denotes a collection of (scalar) fields, $\Phi'$ and $\Phi''$ denote any of its first and second partial derivatives, and $M_\Phi^2$ denotes a diagonalised mass-matrix. (The results in~\cite{klainerman1985} are stated with $M_{\Phi_i}^2\equiv 1$ while those in~\cite{simon1985,simon1993} are stated for $M_{\Phi_i}^2 > 0$. In any case, they hold only for non-vanishing and non-tachyonic masses.) 
For instance, for non-ghostly theories, the general corresponding Lagrangian would read
\begin{align}
    \mathcal{L} = 
    \frac{1}{2}\Phi\left(\Box + M_\Phi^2\right)\Phi
    + \mathcal{O}(\Phi, \Phi', \Phi'')
\end{align}
and the central proof assumption amounts to $\mathcal{O}(\Phi, \Phi', \Phi'')$ starting at cubic order or higher.

In turn, for ghostly theories, the proof, for instance, applies to
\begin{align}    
    \mathcal{L} = 
    \frac{1}{2}\widetilde\Phi\Box^2\widetilde\Phi
    + \frac{1}{2}\widetilde\Phi\alpha\Box\widetilde\Phi
    + \frac{1}{2}\widetilde\Phi\beta\widetilde\Phi
    + \mathcal{O}(\widetilde\Phi, \widetilde\Phi', \widetilde\Phi'')\;,
\end{align}
where the first term is assumed to be diagonal and $\alpha$ and $\beta$ denote matrix-valued parameters. These can be used -- assuming non-degeneracy -- to adjust the mass spectrum of the order-reduced linear problem, see below. The term $\mathcal{O}(\widetilde\Phi, \widetilde\Phi', \widetilde\Phi'')$ denotes, once more, arbitrary nonlinearities starting at cubic order or higher. The associated 4th-order equations of motion can be written as
\begin{align}
    0 &= 
    \Box^2\widetilde\Phi 
    + \alpha\Box\widetilde\Phi 
    + \beta\widetilde\Phi
\end{align}
or in 2nd-order form
\begin{align}
    \Box\widetilde\Phi &= \widetilde\Psi
    \\
    \Box\widetilde\Psi &=
    - \alpha\widetilde\Psi 
    - \beta\widetilde\Phi
    + F(\widetilde\Phi, \widetilde\Phi', \widetilde\Phi'';\widetilde\Psi, \widetilde\Psi', \widetilde\Psi'')
\end{align}
for which one can always diagonalise the linear terms and achieve a non-tachyonic mass-spectrum by suitable choice of $\alpha$ and $\beta$, i.e., we can bring the field equations to a form
\begin{align}
    \Box\Phi_+ &= 
    M_+^2\,\Phi_+
    + F_+(\Phi_+, \Phi_+', \Phi_+'';\Phi_-, \Phi_-', \Phi_-'')
    \;,
    \\
    \Box\Phi_- &= 
    M_-^2\,\Phi_-
    + F_-(\Phi_+, \Phi_+', \Phi_+'';\Phi_-, \Phi_-', \Phi_-'')
    \;.
\end{align}
The diagonalisation of the linear terms does not change the order of the nonlinear terms and, thus, the system is again of the desired form, as long as $\mathcal{O}(\widetilde\Phi, \widetilde\Phi', \widetilde\Phi'')$ is polynomial and starts at cubic order or higher.

We assume that the result can be generalised inductively to capture a subclass of higher-derivative theories of arbitrary order. For all such theories, the theorem then establishes small-data global stability. We will return to this question in future work. 

We also note that the class of gravitational EFTs for which local well-posedness has recently been established~\cite{Figueras:2024bba}, as well as the toy-model EFT in~\cite{Figueras:2025gal}, require more sophisticated methods (see, e.g.~\cite{klainerman1986null}) and small-data global stability remains open. Such EFTs do not fulfil the assumptions of the small-data global stability theorems in~\cite{klainerman1985,simon1985,simon1993} because their 2nd-order field equations contain massless modes and linear contributions with 1st-order derivatives. While these do not obstruct from a proof of local well-posedness, they break the assumptions of the theorems in~\cite{klainerman1985,simon1985,simon1993} such that a proof of small-data global stability remains open.
\\

With this small excursion, we return to limiting our investigation to ghostly systems with polynomial non-derivative interactions, i.e., to the class of systems specified in~\cref{eq:field-theory-Lagrangian}.


\section{Numerical evidence for large-data global stability}
\label{sec:numerical}

We emphasize that the mathematical theorems reviewed in the previous sections hold only for sufficiently small initial data. At the same time, physical conditions on the potential $V(\phi,\chi) = V_{\phi}(\phi) + \sigma V_{\chi}(\chi) + V_\text{int}(\phi,\chi)$ have been identified in the context of integrable point-particle systems~\cite{Deffayet:2023wdg}, with the expectation (and numerical evidence) that they extend also to non-integrable point-particle systems. These conditions may be summarised as the requirement that, at large field-values -- that is outside of a compact region in field-space -- the self-interactions in $V_{\phi} + \sigma V_{\chi}$ are sufficiently dominant in comparison to the interactions in~$V_\text{int}$. Motivated by this, we seek to understand under which conditions small-data global stability results may extend to global stability results that apply to \emph{all} compact initial data. To do so, we perform numerical experiments to (i) quantify the critical initial data region that bounds the small-data global stability results, (ii) identify how self-interactions can quench the onset of instabilities outside this small-data global stability region (iii) and, from this, derive expectations for the physical conditions which ensure global stability for \emph{all} compactly supported initial data.

For concreteness and better numerical efficiency, we focus on spherical symmetry and on a class of radially incoming wave packets with a characteristic amplitude $A$ and characteristic width (or equivalently inverse frequency) $w$. 
Within this setup, the incoming wave packets will interact in a central interaction region around the origin. They will then either lead to a (finite-time) divergence/runaway or scatter and form an outgoing wave packet. We interpret the former case as a counter-example to global stability. 

Further details of this numerical setup are given in~\cref{sec:scattering} below where we also show that the respective parameter space of the problem is fully characterised by three dimensionless ratios. The following sections
present the conclusions that we can draw from the respective numerical solutions.

\subsection{Numerical setup: Spherically-symmetric scattering}
\label{sec:scattering}

In the following, we provide all necessary details to obtain the numerical solutions. 

\subsubsection{Reduction to spherical symmetry}

We impose spherical symmetry in space, denoting with $r$, the respective radial coordinate. The semi-linear Klein-Gordon system in~\cref{eq:eoms} can then be rewritten in first-order form as
\begin{align}
    \partial_t\phi &\equiv 
    \hat{\phi}
    \;,
    \notag\\*
    \partial_t\hat{\phi} &=
    - m_\phi^2\,\phi
    + \Delta_r\,\phi
    - \partial_\phi V
    \;,
    \notag\\*
    \partial_t\chi &\equiv 
    \hat{\chi}
    \;,
    \notag\\*
    \partial_t\hat{\chi} &=
    - m_\chi^2\,\chi
    + \Delta_r\,\chi
    - \sigma\,\partial_{\chi} V
    \;,
\label{eq:first_order}
\end{align}
where $\Delta_r=\partial_r^2 + \frac{N-1}{r}\,\partial_r$ denotes the $N$-dimensional spatial Laplacian in spherical symmetry.
Initial data is thus given by four initial conditions $\{\phi,\,\hat\phi,\,\chi,\,\hat\chi\}$, corresponding to two physical degrees of freedom.
We use fourth-order central finite differencing kernels to numerically solve the initial value problem with (an updated version of) the \texttt{julia} code\footnote{
    The complete \texttt{julia} code is available on \href{https://github.com/aaron-hd/ghostlyPDE_spherical}{\texttt{github.com/aaron-hd/ghostlyPDE\_spherical}}.
} introduced in~\cite{Deffayet:2025lnj}. 

\subsubsection{Physical domain and boundary conditions}

We work in a scattering domain of radius $r=R$ which implicitly sets our units. This means that all dimensionful physical parameters are measured (in natural units) of the length $R$ and can thus be generalised to spheres of arbitrary radius by appropriate rescaling. As detailed below, our initial data corresponds to physical scattering from $T=-R$ to $T=R$. In practice, we solve the equations in a physical domain $L>R$ of suitably larger radius such that the initial data is effectively compactly supported within a sphere of $L$. More specifically, the radial initial-data profile falls of sufficiently fast such that at $r>L$ its value is negligible in comparison to the numerical precision. Thus, the assumptions of small-data global stability theorems apply. 
At the inner boundary, we enforce regularity conditions\footnote{The author wishes to thank Ramiro Cayuso for helpful communication concerning the numerical implementation of regularity conditions in radial evolution codes.}.
At the outer boundary, we implement open boundary conditions. The latter allows us to work in a spatial domain of $L=2R$, while nevertheless investigating late-time behaviour within the central scattering region $r<R$ up to times $T\gg R$. Alternatively, the same results can be obtained, albeit with larger numerical effort, in a sufficiently large spatial domain in which the scattered wave-packets never reach the boundary.

\subsubsection{Initial data}
\label{sec:initial-data}

For definiteness, we focus on incoming wave packets with a radial Gaussian profile, i.e.,
\begin{align}
    \phi_0(r) &=
    \frac{A_\phi}{r^{(N-1)/2}}\,
    \exp\left(
        -\frac{\left(r-R\right)^{2}}{2\,w_\phi^{2}}
    \right)
    \;,
    \notag\\
    \hat\phi_0(r) &=
    \frac{A_\phi}{r^{(N-1)/2}}\,
    \frac{r-R}{w_\phi^{2}}
    \exp\left(
        -\frac{\left(r-R\right)^{2}}{2\,w_\phi^{2}}
    \right)
    \;,
\end{align}
and equivalently for $\chi$.

In a first setup, we reduce to a one-parameter initial-data family as follows:
First, we fix $w_\phi/R = w_\chi/R = 0.1$. The results of~\cite{Deffayet:2025lnj}, suggest that higher-frequency initial data is more stable not less stable. Based on this, we expect that, for decreasing, $w/R<0.1$, the stability properties will improve.
Second, we identify $A_\phi = \pm|A|$, $A_\chi = \pm|A|$. The latter implies that we perform our numerical analysis with all four sign combinations.
Overall, we have thereby constructed a suitable one-parameter family of initial data with characteristic dimensionless amplitude $\bar{A}\equiv A\times R^{(N-1)/2}$.
We will refer to this as our \emph{``one-parameter family of initial data''} in the following, sometimes adding \emph{``equal-amplitude''} to distinguish form the second setup below.

In a second setup, we pick random $w_\phi/R\in[0.01,0.1]$ and $w_\chi/R\in[0.01,0.1]$ as well as $\bar{a}_\phi\in[0,1]$, and $\bar{a}_\chi\in[0,1]$, all following flat distributions. From the latter, we construct $A_\phi = \bar{a}_\phi\times|\bar{A}|/R^{(N-1)/2}$ and $A_\chi = \bar{a}_\phi\times|\bar{A}|/R^{(N-1)/2}$ such that the dimensionless initial-data amplitudes are distributed around a characteristic amplitude $\bar{A}$, which, again, is our only free parameter. We will refer to this as our \emph{``randomised one-parameter family of initial data''.}

\subsubsection{Dimensional scaling and characteristic dimensionless parameters}
\label{sec:dimensional-analysis}

As discussed above, we set units by the scattering radius $R$ and focus on a family of initial data characterised by a single dimensionless amplitude $\bar{A}\equiv A\times R^{(N-1)/2}$, quantifying the smallness of the initial data.

Similarly, we work with the potentials in~\cref{eq:Vphi,eq:Vchi,eq:Vcross} which are fully determined by the couplings $\lambda_{\ell}$ and $\lambda_{mn}$, as well as the positive integer exponents $\ell,m,n\in\mathbb{N}$. Dimensionless analysis reveals the respective characteristic dimensionless ratios
\begin{align}
    \bar{\lambda}_{\ell} &= 
    \frac{\lambda_{\ell}}{R^{(N+1)-(N-1)\,\ell/2}}
    \\
    \bar{\lambda}_{mn} &=    
    \frac{\lambda_{mn}}{R^{(N+1)-(N-1)\,(m+n)/2}}
\end{align}
depending on the integer exponents $\ell,m,n\in\mathbb{N}$ and the spacetime dimension $N\in\mathbb{N}$.
For simplicity, we will focus on the massless case ($m_\phi=m_\chi=0$) below. Due to dissipation (see~\cref{sec:analytical}), we expect the massive case to be more stable.

\subsection{Scattering solutions for ghostly field theories}
\label{sec:numerical-results:ghostly}

In this section, we present a study of spherically-symmetric scattering solutions in the ghostly case ($\sigma=-1$). First, we consider unquenched ghostly interactions, i.e., we set $\bar{\lambda}_{\ell}=0$: In~\cref{sec:numerical-results:ghostly:small-data}, we verify small-data global stability and delineate the respective threshold. Beyond the threshold, we find (in agreement with~\cite{Deffayet:2025lnj}) that the ghost instability can manifest either as a benign runaway or as a catastrophic finite-time blow up.
Second, we demonstrate how self-interactions can quench the above instability: In~\cref{sec:numerical-results:ghostly:large-data}, we present a systematic study of large-data scattering solutions for the exponents $(N,\ell,m,n)$ to confirm that the above instability can be quenched by sufficiently
dominant self-interactions, i.e., if $\ell>n+m$. For $\ell=n+m$, we find indications for benign behaviour but defer a more detailed study to future work.

\subsubsection{Small-data global stability for unquenched ghostly interactions}
\label{sec:numerical-results:ghostly:small-data}

%
\begin{figure}
    \begin{centering}
        \includegraphics[width=\linewidth, trim={8cm 2.6cm 6cm 0cm}, clip]{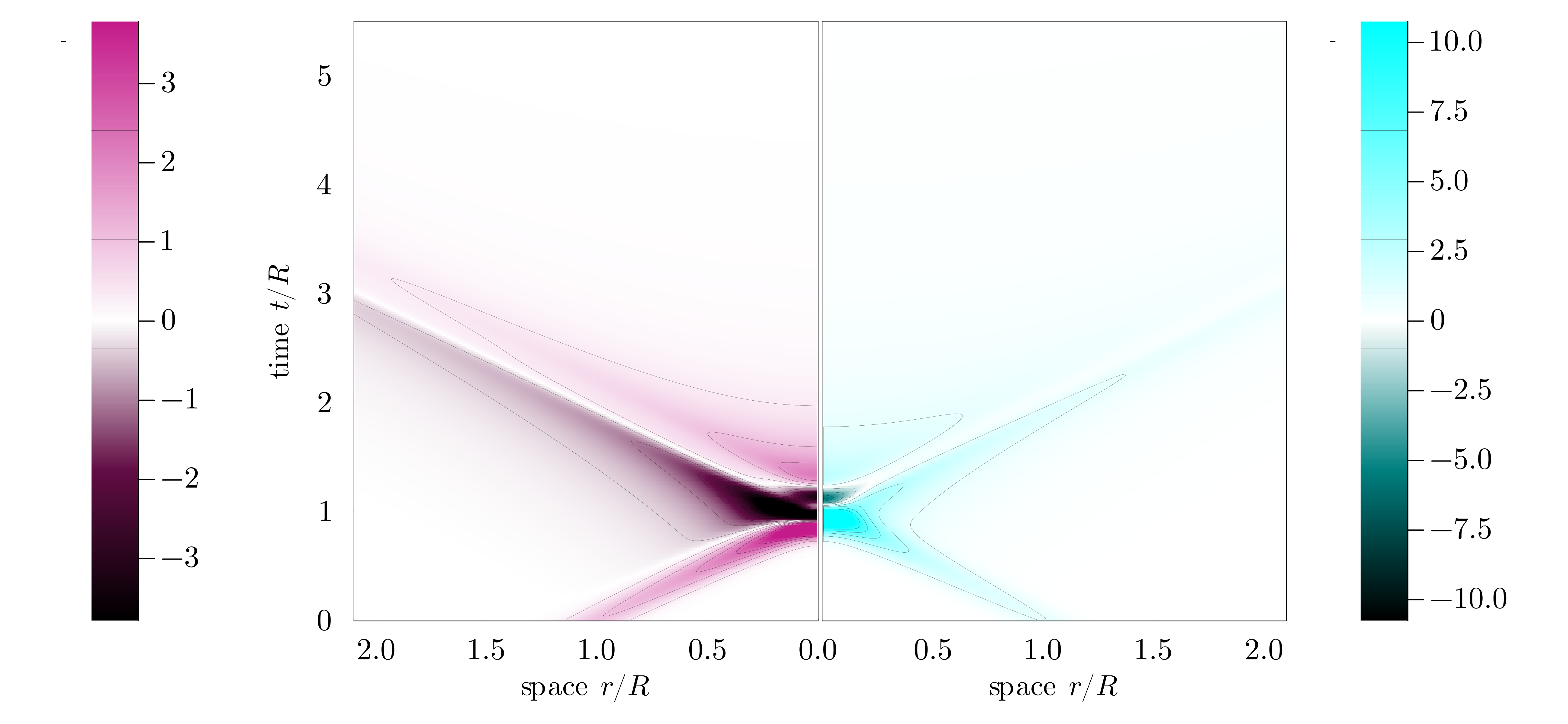}
        \\        
        \includegraphics[width=\linewidth, trim={8cm 2.6cm 6cm 0cm}, clip]{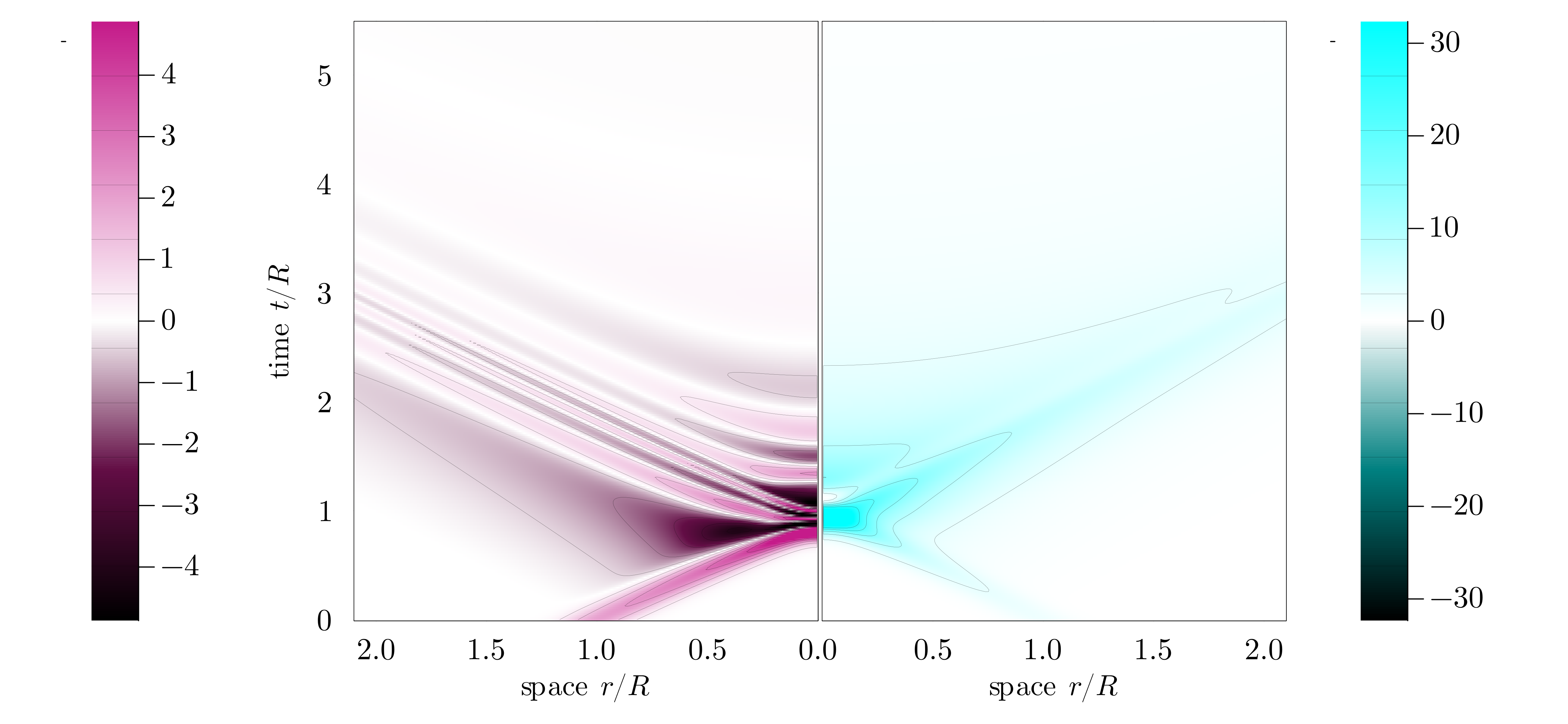}
    \end{centering}
    \caption{\label{fig:example-scattering-unstable-ghostly-benign}
    Exemplary scattering solutions of the characteristic initial-data family (see~\cref{sec:initial-data}) interacting via an unquenched ghostly interaction ($\sigma=-1$), i.e., $V(\phi, \chi)=\lambda_{22}\,\phi^2\,\chi^2$ with $\lambda_{22}=1$, in $N=3$ spatial dimensions. In the upper (lower) panel, $|A|=1$ ($|A|=2$). In both panels, we pick equal and positive sign initial data. The left and right portions of the plots respectively show the radial profile of $\phi$ and $\chi$. 
    \href{https://zenodo.org/records/17178254}{Animations available online.}
    }
\end{figure}
\begin{figure}
    \begin{centering}
        \includegraphics[width=\linewidth, trim={3cm 1cm 3cm 1cm}, clip]{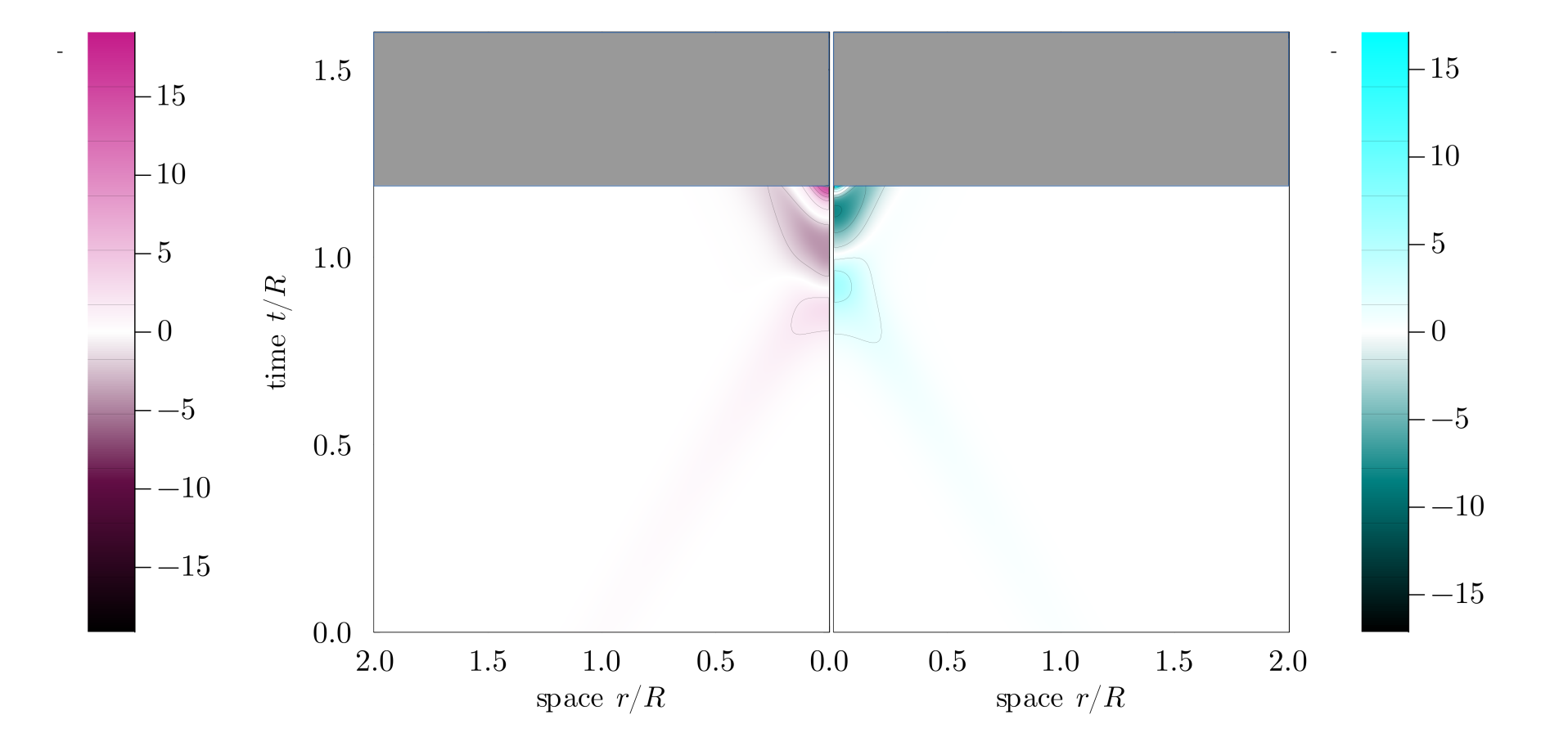}
    \end{centering}
    \caption{\label{fig:example-scattering-unstable-ghostly-catastrophic}
    Exemplary scattering solution of the characteristic initial-data family (see~\cref{sec:initial-data}) interacting via an unquenched ghostly interaction ($\sigma=-1$), i.e., $V(\phi, \chi)=\lambda_{33}\,\phi^3\,\chi^3$ with $\lambda_{33}=1$, in $N=3$ spatial dimensions. While at small-data global stability holds for amplitudes $|A|\lesssim0.3$, we show a supercritical case with $|A|=0.31$, for which the solution blows up at finite time due to a divergence in the central scattering solution.
    The left and right portions of the plots respectively show the radial profile of $\phi$ and $\chi$. 
    \href{https://zenodo.org/records/17178254}{Animations available online.}
    }
\end{figure}
\begin{figure}
    \begin{centering}
        \hfill
        \includegraphics[width=0.7\linewidth]{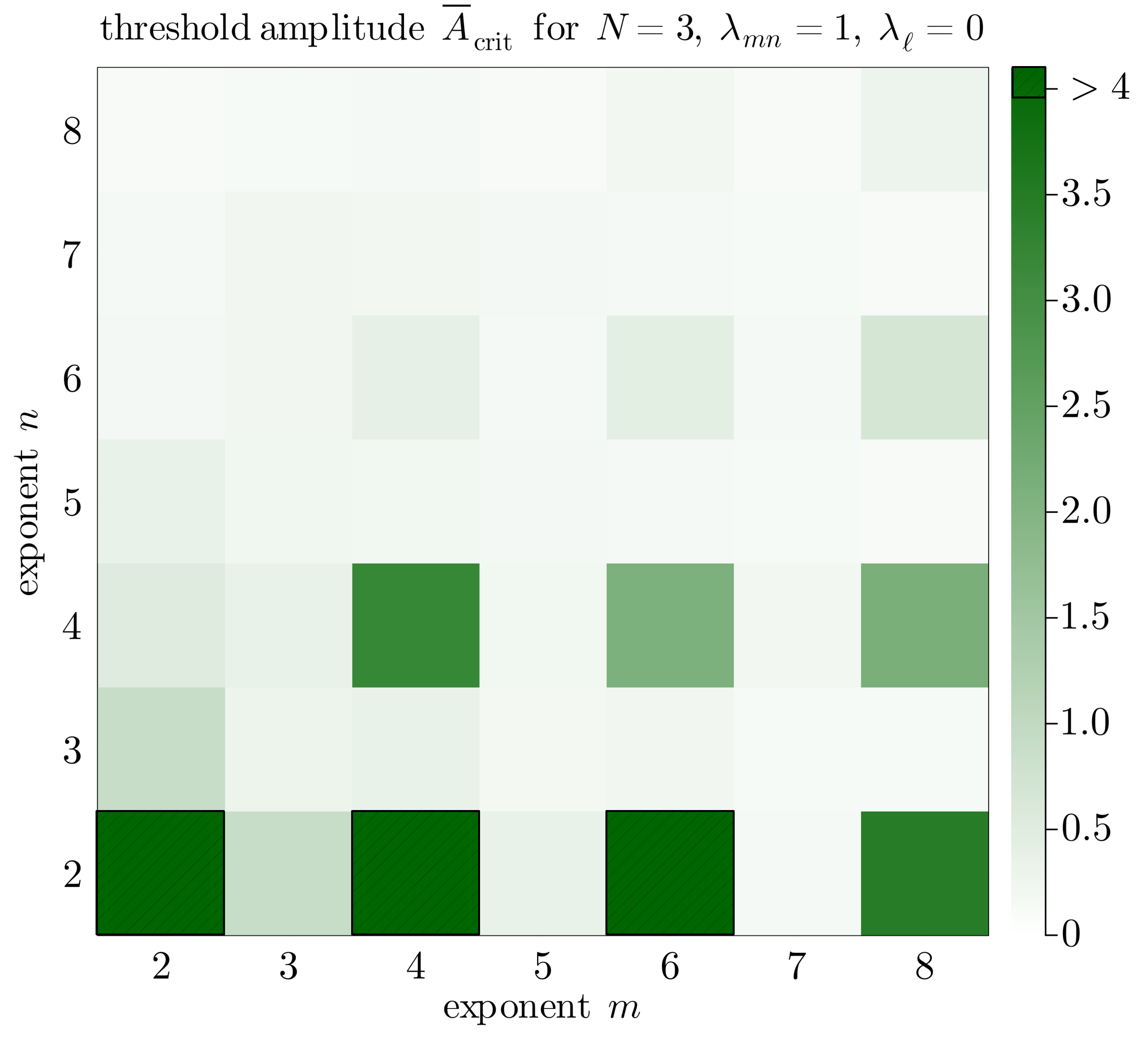} 
        \hfill
    \end{centering}
    \caption{\label{fig:critical_amplitude_ghostly_unquenched}
    Apparent critical threshold $\bar{A}_\text{crit}$ (see colour legend) in a one-parameter initial data family (see~\cref{sec:initial-data}) for spherically symmetric scattering in $(3+1)$ dimensional ghostly field theories without self-interactions (i.e., $\lambda_\ell=0$) as a function of the interaction exponents $m, n\in\mathbb{N}$. Above the apparent critical threshold, we observe a finite-time divergence before  the critical threshold, scattering occurs; above the critical threshold. Hatched tiles indicate that no critical threshold could be found.
    }
\end{figure}

Here, we consider unquenched ghostly interaction potentials, hence, we set $\bar{\lambda}_{\ell}=0$, pick $\sigma=-1$, and, for concreteness, we set $\bar{\lambda}_{mn}=+1$. The case $\bar{\lambda}_{mn}=-1$ follows from the exchange of $\phi\leftrightarrow\chi$, i.e., from the exchange of $m\leftrightarrow n$.
As a preface, we note that, of course, all investigated examples confirm the proven theorem on small-data global stability.

With the extension from small-data global stability to large-data global behaviour in mind, we investigate the behaviour of the scattering solution as a function of growing characteristic amplitude $\bar{A}$. We start by discussing the possible types of behaviour in exemplary cases. 

The first type of behaviour occurs if the unquenched ghostly interactions lead to a benign runaway instability. Such benign behaviour has been previously observed, even in (1+1) dimensions~\cite{Deffayet:2025lnj}. Scattering solutions for the exemplary case $(N,m,n)=(3,2,2)$ are shown in~\cref{fig:example-scattering-unstable-ghostly-benign}. The respective point-particle system is unstable, even for arbitrarily small data. However, the field theory exhibits small-data global stability due to dominance of geometric spreading into the spatial dimensions, see~\cref{sec:analytical}. In fact, for $n=m=2$, the ghost instability is apparently benign, as has previously been observed in $(1+1)$ dimensions~\cite{Deffayet:2025lnj}: As the initial amplitude $\bar{A}$ increases, the component energies of each individual field (see~\cref{eq:component-energies}) seem to grow without bound but the numerical solutions can apparently be extended to arbitrarily large times. In practice, this requires higher and higher resolution of the numerical simulation. Nevertheless, this suggests global existence for all compact initial data with arbitrarily large amplitude.
As discussed in~\cite{Deffayet:2025lnj}, a physical explanation can be given in terms of effective masses.
To do so, we consider the explicit 2nd-order field equations which can be written as
\begin{alignat}{4}
	\Box\,\phi 
    &=&
    \left(
        m_\phi^2 + m\,\lambda_{mn}\,\phi^{m-2}\,\chi^{n}
    \right)\phi
    &\;\equiv\;& 
    m_{\phi,\text{eff}}^2\,\phi
    \;,
	\notag\\
	\Box\,\chi 
    &=&
    \left(
        m_\chi^2 + m\,\lambda_{mn}\,\phi^{m}\,\chi^{n-2}
    \right)\chi
    &\;\equiv\;& 
    m_{\chi,\text{eff}}^2\,\chi
    \;,
	\label{eq:eoms}
\end{alignat}
where we have identified the effective squared masses $m_{\phi,\text{eff}}(\phi,\chi)^2$ and $m_{\chi,\text{eff}}(\phi,\chi)^2$. For $\lambda_{mn}>0$ and even $m,\,n$ the effective mass of $\phi$ cannot become locally tachyonic, i.e., $m_{\phi,\text{eff}}^2>0$. In~\cite{Deffayet:2025lnj}, this has been connected to the possibility of benign ghostly interactions, i.e., field theories without finite-time divergences.

The second type of behaviour occurs if the unquenched ghostly interaction can become sufficiently focusing as to overcome the effects of geometric spreading and thus lead to a catastrophic instability at finite time. Scattering solutions for the exemplary case $(N,m,n)=(3,3,3)$ are shown in~\cref{fig:example-scattering-unstable-ghostly-catastrophic}. 
In contrast to the benign behaviour discussed above, the catastrophic case exhibits a sharp threshold amplitude $\bar{A}_\text{crit}$: For $\bar{A}<\bar{A}_\text{crit}$, we confirm small-data global stability (see~\cref{sec:analytical}), scattering still occurs, the evolution can be continued to arbitrarily late time, and all component energies remain finite. However, for $\bar{A}>\bar{A}_\text{crit}$, we find a catastrophic instability, i.e., and apparent blow up of the solution in finite time. 
As established by mathematical proof (see~\cref{sec:analytical}), and in contrast to the respective point particle system, there still exists a small-data global stability region at sufficiently small amplitude $\bar{A}$. 
\\

After demonstrating the two different types of behaviour, we obtain the dimensionless threshold amplitude $\bar{A}_\text{crit}(N,m,n)$ that delimits the small-data global stability region. Our main results are summarised in~\cref{fig:critical_amplitude_ghostly_unquenched} which provides a visual summary of small-data global stability and quantifies the respective threshold amplitude. The figure is obtained by bisecting for $\bar{A}_\text{crit}(N,m,n)$, for each set of $N,m,n\in\mathbb{N}$ and projecting onto two-dimensional hypersurfaces for which we present a discrete heat map of $\bar{A}_\text{crit}(N,m,n)$.
To be precise, we evolve the evolution up to $T/R=10$ (or until blow up occurs) and bisect in an interval $\bar{A}\in[0,10]$. We find that, if a finite $\bar{A}_\text{crit}(N,m,n)$ exists, it always lies below $\bar{A}<2$. Otherwise, the bisection converges to its upper interval boundary from which we conclude that no finite threshold amplitude exists. In~\cref{fig:critical_amplitude_ghostly_unquenched}, we indicate the latter cases as hatched heat map tiles.
\\

We close this section with the following overall conclusions:
At small data, all our numerical solutions confirm that there exists an open region of initial data in which global stability of the vacuum leads to global existence and scattering.
At large-data, we either find benign or catastrophic runaway behaviour, depending on the specific type of nonlinear interaction.

All of the above also holds for single-field interactions in an unstable (focusing) potential. This suggests, once more (see also~\cite{Deffayet:2025lnj}), that the ghost instability -- at least when mediated by non-derivative interactions -- is not different from other potential instabilities.
We will further substantiate this overall picture in the next section, in which we demonstrate that sufficiently dominant self-interactions can apparently quench the ghost instability.

\subsubsection{Apparent large-data global stability for quenched ghostly interactions}
\label{sec:numerical-results:ghostly:large-data}

%
\begin{figure}
    \centering
    \includegraphics[width=0.95\linewidth, trim={8cm 2.6cm 6cm 0cm}, clip]{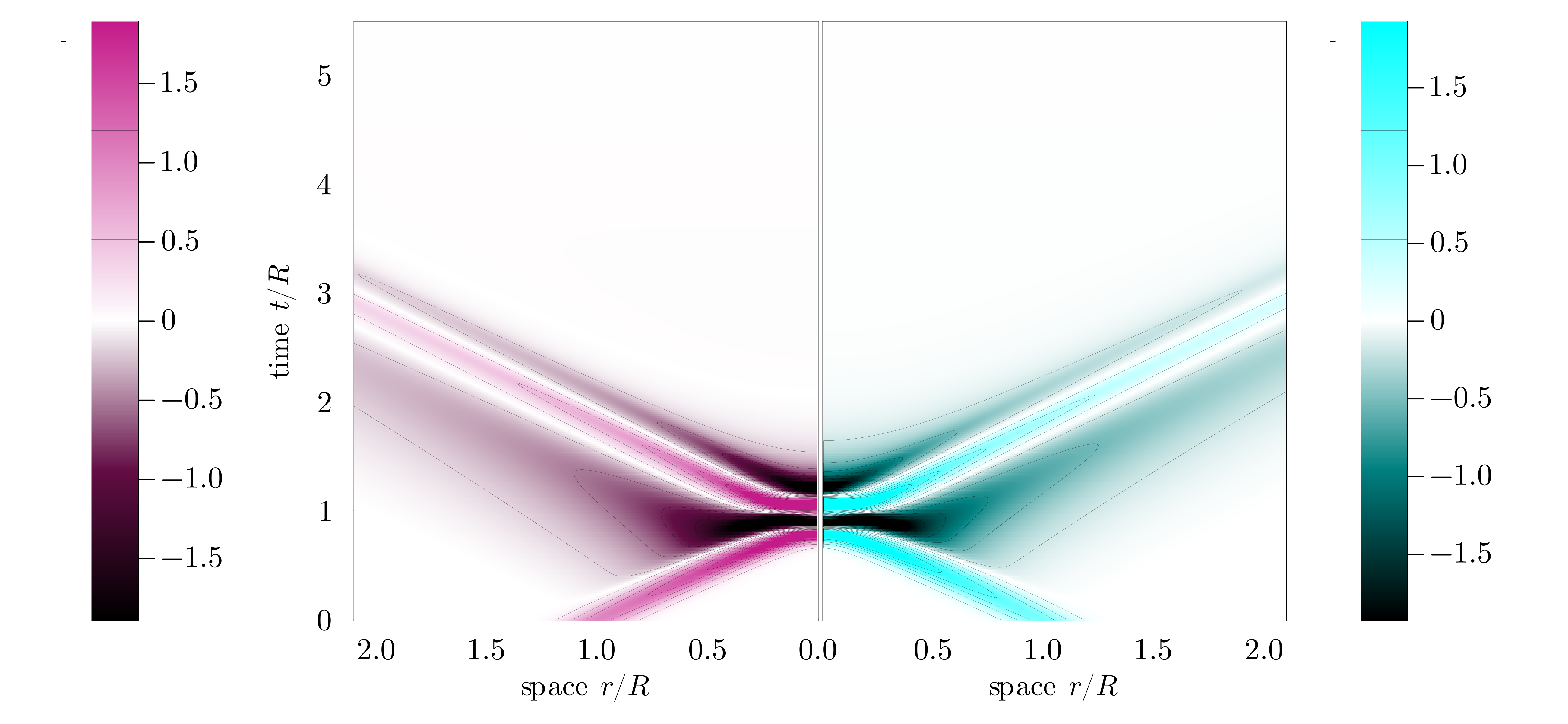}
    \\
    \includegraphics[width=0.95\linewidth, trim={8cm 2.6cm 6cm 0cm}, clip]{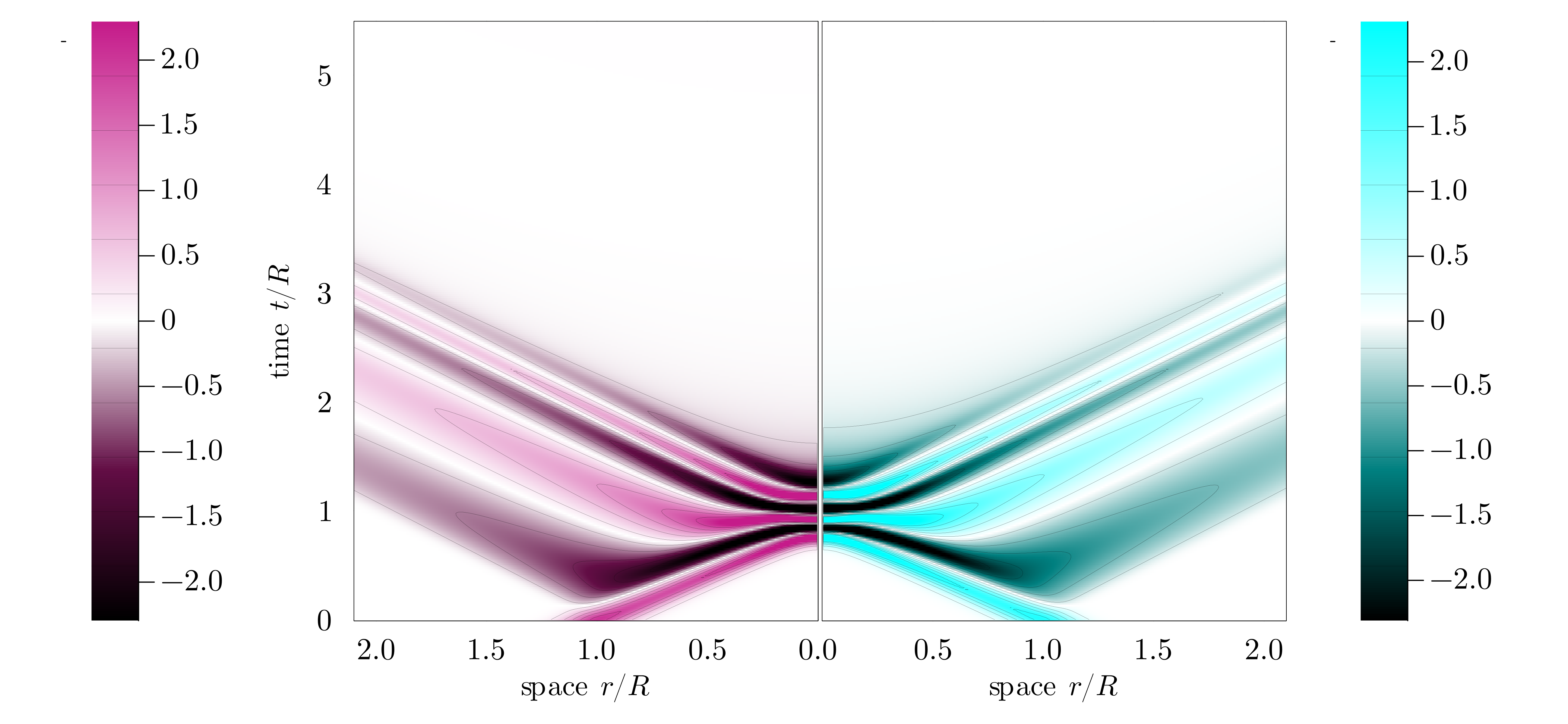}
    \\
    \includegraphics[width=0.95\linewidth, trim={8cm 2.6cm 6cm 0cm}, clip]{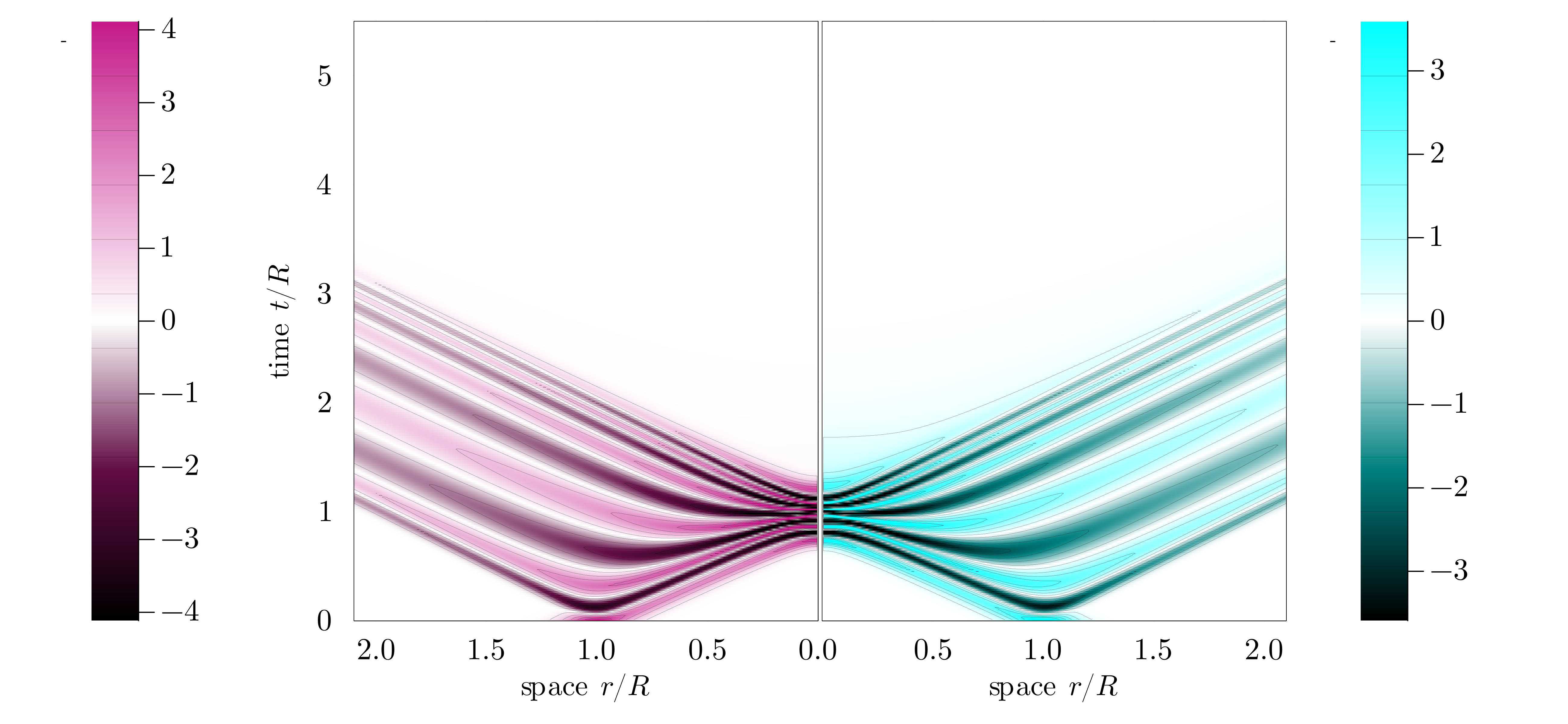}
    \caption{\label{fig:examplary-3622}
    Exemplary series of scattering solutions with increasing characteristic amplitude ($|A|=1,\,2,\,4$ from top to bottom) for the case $(N,\ell,m,n)=(3,6,2,2)$. We pick equal and positive sign initial data but the other sign choices exhibit the same behaviour. The left and right portions of the plots respectively show the radial profile of $\phi$ and $\chi$.
    \href{https://zenodo.org/records/17178254}{Animations available online.}
    }
\end{figure}

Here, we investigate whether small-data global stability can be extended to large-data global stability if the ghost instability is quenched by self-interactions. 
From the point-particle analysis in~\cite{Deffayet:2023wdg}, we expect that ghostly theories ($\sigma=-1$) can be stable if self-interactions are \emph{sufficiently dominant} in comparison to ghostly interactions that mediate between the opposite-sign kinetic terms, see also~\cref{sec:introduction}. For the class of classical field theories with potential interactions under investigation, see~\cref{eq:field-theory-Lagrangian}, we expect that it is sufficient to demand that the self interactions are stable (defocusing) and of higher polynomial degree than the ghostly interactions, i.e.,  
\begin{align}
    \ell>n+m\;.
\end{align}
Due to small-data global stability (see~\cref{sec:analytical} for proof and the previous \cref{sec:numerical-results:ghostly:small-data} for demonstration), we moreover expect that geometric spreading leads to improved stability of the field theory in comparison to the respective point-particle system.
Here, we verify these expectations for spherically-symmetric scattering solutions.

First, we demonstrate the above expectations for the case of $(N,\ell,m,n)=(3,6,2,2)$: In~\cref{fig:examplary-3622}, we successively increase the characteristic amplitude $\bar{A}$ of our one-parameter family of initial data. As we increase the characteristic amplitude $\bar{A}$, we observe no sign of instability. All solutions scatter and decay back to the trivial vacuum. In fact, we can observe that, for large amplitudes, the ghostly interaction indeed becomes an increasingly irrelevant perturbation and the scattering solution approximates the respective decoupled solution (with $\lambda_{mn}=0$) increasingly well.
\\

\begin{figure}
    \centering
        \includegraphics[width=0.55\linewidth]{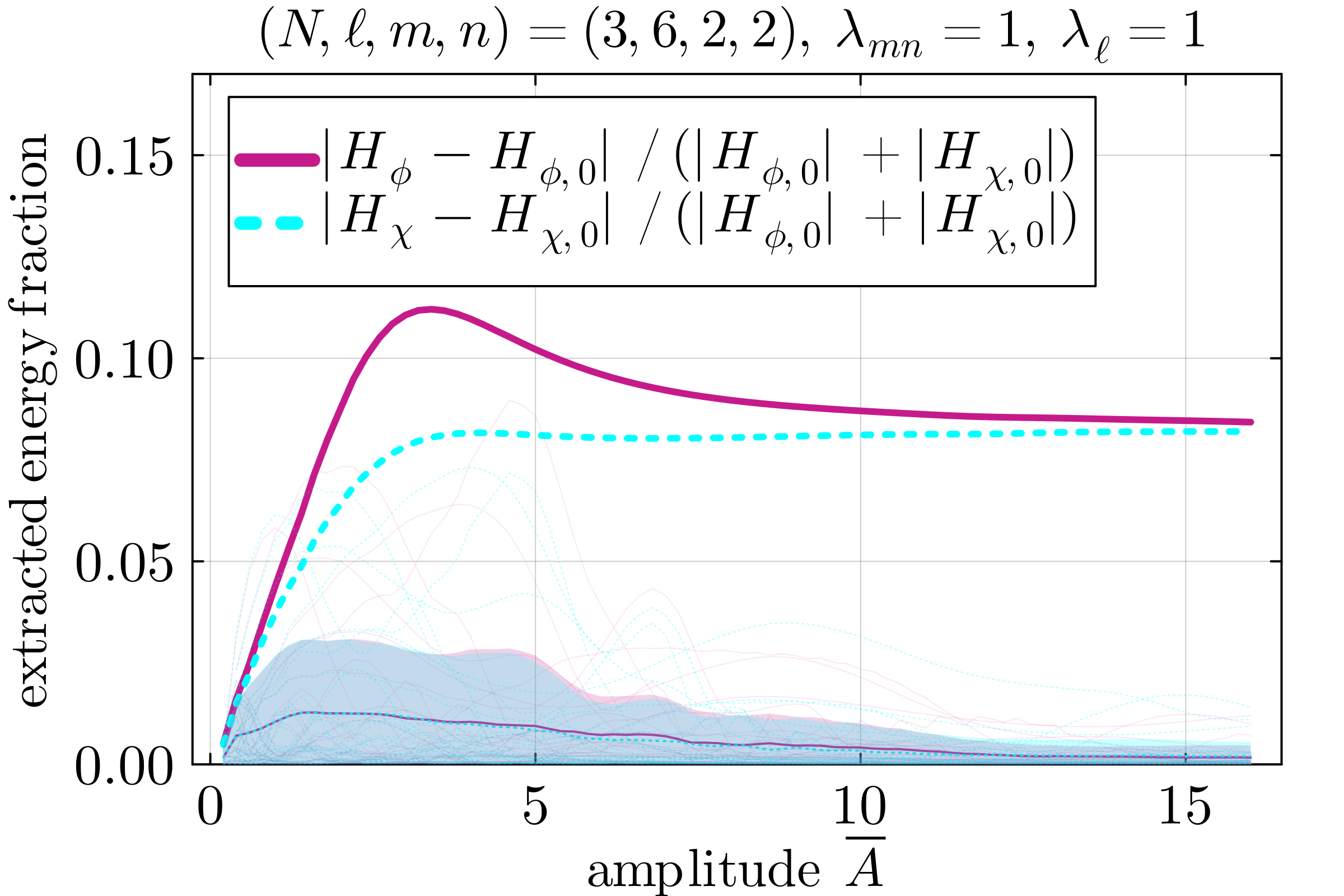} 
    \caption{\label{fig:relative-extracted-energy}
    We demonstrate (for the exemplary case $(N,\ell,m,n)=(3,6,2,2)$) that the extracted energy within each individual field remains bounded as a function of growing characteristic amplitude. This holds, both, for the equal-amplitude (here with positive and equal signs) one-parameter family of initial data (thick lines) and for the randomised one-parameter family of initial data, for which we show the average of 30 random cases (thin lines) and the respective statistical error (shaded regions).
    }
\end{figure}

To further substantiate evidence for the apparent large-data global stability, we investigate how much energy can be extracted by the individual fields in the respective scattering process. To do so, we track the total Hamiltonian $H$ and the individual component energies
\begin{align}
    H_\phi(T) &= 
    \int_{\mathbb{R}^N}\left[
        \frac{1}{2}\left(\dot{\phi}^2 + \phi'^2 + m_\phi^2\phi^2\right)
        + V_\phi[\phi]
    \right]
    d^Nx
    \;,
    \label{eq:component-energies}
\end{align}
and equivalently for $\chi$, see also~\cite{Deffayet:2023wdg}. We compare the increase in $H_\phi(T)$ (as well as $H_\chi$) to their initial values $H_{\phi/\chi,0}=H_{\phi/\chi}(T=0)$, i.e., we determine $|H_\phi-H_{\phi,0}|/|H_{\phi,0}+H_{\chi,0}|$ and $|H_\phi-H_{\phi,0}|/|H_{\phi,0}+H_{\chi,0}|$ at asymptotically late time $T\gg 1$. We obtain this measure for the extracted energies of the individual fields, as a function of increasing characteristic initial amplitude $\bar{A}$. The exemplary case of $(N,\ell,m,n)=(3,6,2,2)$ is shown in~\cref{fig:relative-extracted-energy}. 
We note that the symmetry between $\phi$ and $\chi$ is broken by the sign of $\lambda_{22}=+1$.
These results clearly show that the relative extracted energies of both fields remain bounded, further demonstrating the expectation of global stability for all initial data.

We emphasise that~\cref{fig:relative-extracted-energy} contains results, both for the equal-amplitude one-parameter family of initial data (thick lines), as well as, for an averaged set of instances in the random one-parameter family of initial data (thin lines). It thus also suggests that large-data global stability seems to hold for any generic choice of spherically-symmetric initial data.
At the same time, we caution that this does not exclude the possibility of fine-tuned blow-up solutions and that a rigorous proof of large-data global stability remains outstanding.
\\

\begin{figure}
    \begin{centering}
        \hfill
        \includegraphics[width=0.48\linewidth]{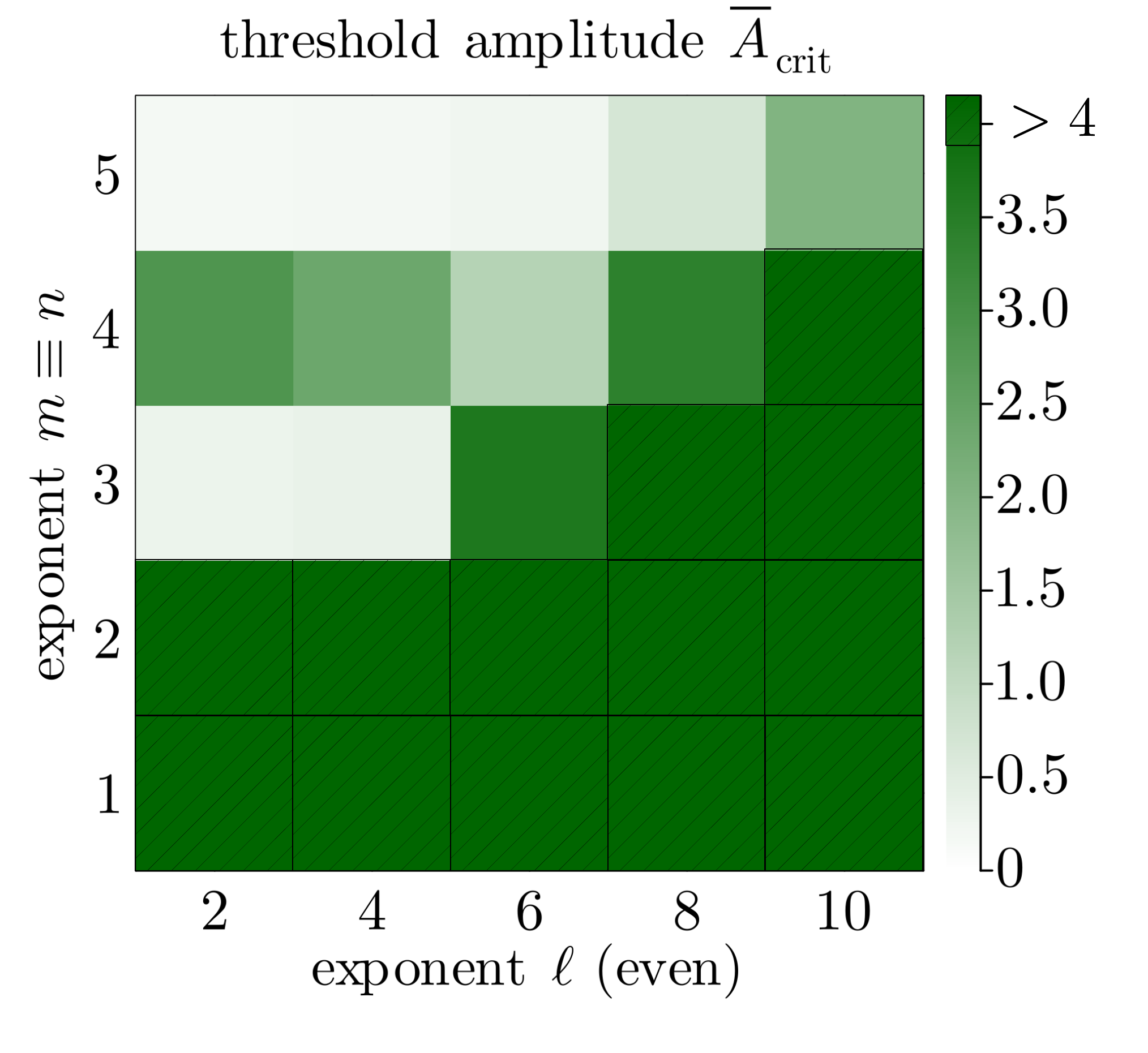} 
        \includegraphics[width=0.48\linewidth]{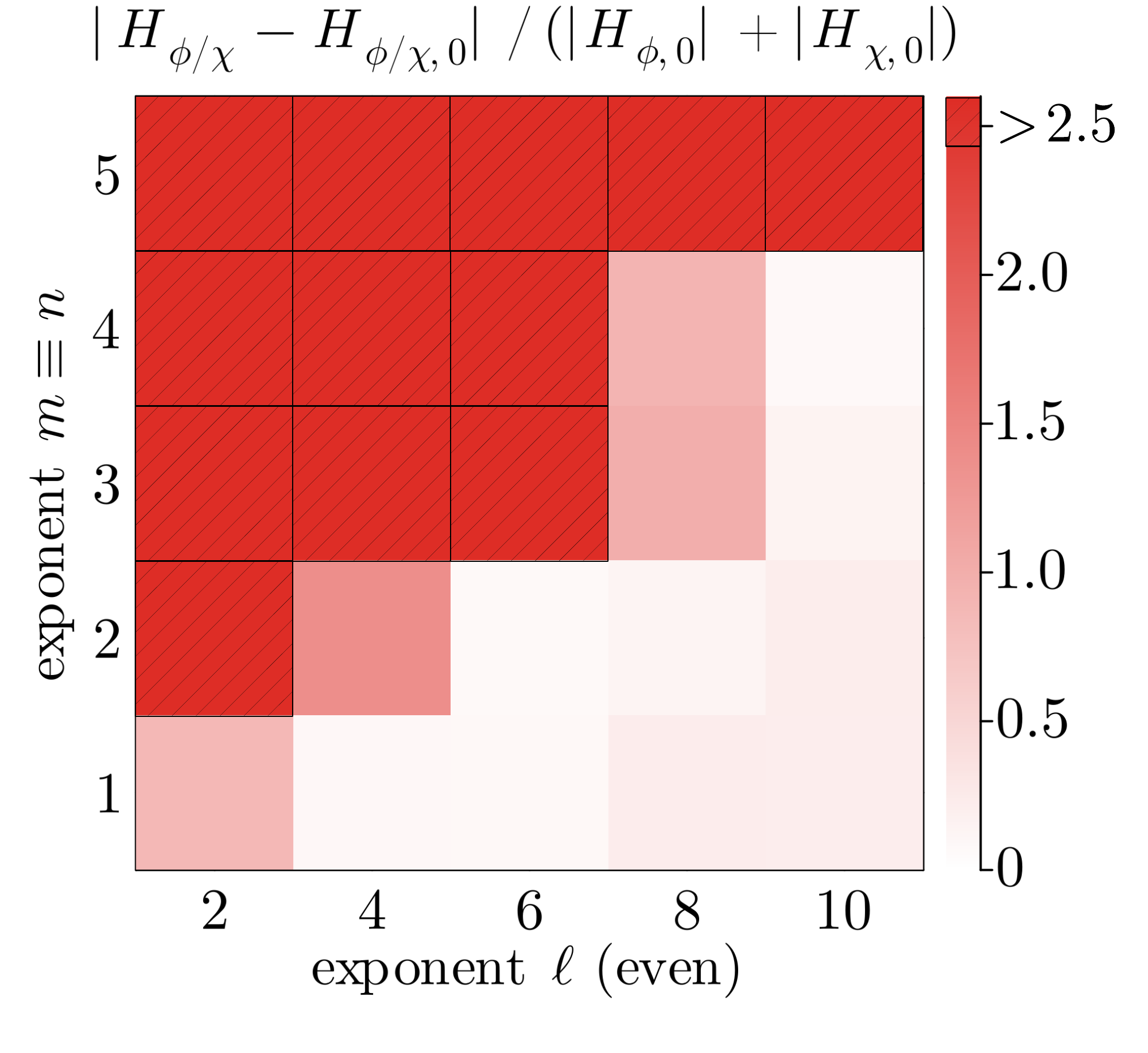} 
        \hfill
    \end{centering}
    \caption{\label{fig:critical_amplitude_ghostly_quenched}
    Threshold amplitude and corresponding extracted energy fraction (slightly below threshold) for spherically symmetric scattering in $(3+1)$ dimensional ghostly field theories ($\sigma=-1$) with self-interaction coupling $\bar{\lambda}_{\ell}=1$ and ghostly interaction coupling $\bar{\lambda}_{mn}=1$ as a function of the exponents $\ell\in\mathbb{N}$ and $m\equiv n\in\mathbb{N}$. Results are obtained for the one-parameter family of initial data (see~\cref{sec:initial-data}) and show the minimal threshold amplitude / maximal energy extraction among all four sign combinations. Hatched tiles on the left indicate that no threshold could be found. Hatched tiles on the right indicate that the extracted energy fraction seems to grow without bound.
    }
\end{figure}

Finally, we verify that the same results hold for $N=3$ and across all integer combinations in the ranges $0 < m,n\leqslant 5$ and even $2 < \ell\leqslant 10$.  
A summary of our results is presented in~\cref{fig:critical_amplitude_ghostly_quenched}. The left panel shows that for $\ell\geqslant n+m$, we do not find any threshold to small-data global stability, suggesting that global existence of the solution extends to arbitrarily large data.
The right panel quantifies the maximal extraced energy as demonstrated for $(N,\ell,m,n)=(3,6,2,2)$ in~\cref{fig:relative-extracted-energy}. In both panels of~\cref{fig:critical_amplitude_ghostly_quenched}, we focus on the non-randomised one-parameter family of initial data and have obtained each heatmap tile by bisection within the initial amplitude range $\bar{A}\in[0,10]$.
This visualisation quantifies the following conclusion
\begin{itemize}
    \item If we can identify a finite threshold amplitude $\bar{A}_\text{crit}$, this implies that initial data with $\bar{A}>\bar{A}_\text{crit}$ leads to finite time divergence. Naturally, at the divergence, also the energy fractions diverge. The respective solution provides a numerical counterexample to large-data global existence. This occurs for all cases with non-hatched tiles in the left panel of~\cref{fig:critical_amplitude_ghostly_quenched}.
    \item If we cannot identify any threshold amplitude but nevertheless observe that the extracted energy fraction grows without bound as a function of growing amplitude, we interpret this as a sign of benign runaway behaviour, i.e., large-data global existence, but not necessarily large-data global stability. This occurs for all cases with hatched tiles in both panels of~\cref{fig:critical_amplitude_ghostly_quenched}.
    \item If there is no identifiable threshold amplitude and the extracted energy fraction remains finite, we interpret this as numerical evidence, not only for large-data global existence, but also for large-data global stability. This occurs for all cases with non-hatched tiles in the right panel of~\cref{fig:critical_amplitude_ghostly_quenched}.
\end{itemize}
All of the above, of course, falls short of a conclusive proof of large-data global stability, and, especially for high (supercritical) exponents, it is numerically challenging to distinguish between benign and blow-up behaviour. Nevertheless, our results support the expectation that small-data global stability can be extended to large-data global stability if the ghost instability is quenched by self-interactions.

%

\section{Physical expectations for the stability of ghosts in classical field theory}
\label{sec:expectations}

Looking at (i) small-data global stability proofs (see~\cref{sec:analytical}), (ii) the large-data numerical scattering solutions in spherical symmetry (see~\cref{sec:numerical}) and (iii) apparent agreement with physical insights on point-particle systems~\cite{Deffayet:2023wdg}, a consistent picture of global stability emerges: Ghostly interactions between opposite-sign kinetic terms can be quenched by sufficiently dominant self-interactions. We find the above combination of insights sufficiently convincing to voice the following expectation, at least in spherical symmetry.
\begin{quote}
    Ghostly field theories, in which two fields $\phi$ and $\chi$ with opposite-sign kinetic terms and non-derivative self-interactions $V_\phi$ and $V_\chi$ interact by a non-derivative potential $V_\text{int}(\phi,\,\chi)$, i.e., systems as in~\cref{eq:field-theory-Lagrangian}, exhibit global stability for \emph{all} compactly supported initial data if
    (i) small-data global stability holds and (ii)
    $V_\text{int}$ is
    overcome at large-field values (i.e., outside of a compact region in field space) by sufficiently dominant stable self-interactions in $V_\phi$ and $V_\chi$, for which equivalent global stability results of the decoupled theory have been (or can be) established.
\end{quote}
For the polynomial interactions in~\cref{eq:Vphi,eq:Vchi,eq:Vcross}, we can give precise meaning to the term ``sufficiently dominant''. For large-data global stability in this class of polynomial ghostly theories, it seems sufficient that $V_\text{int}$ is of lower polynomial degree than, both, $V_\phi$ and $V_\chi$, just like for the respective point particle models~\cite{Deffayet:2023wdg}. 
Hence, while small-data global stability places a lower bound on the degree of $V_\text{int}$, large-data global stability seems to place an upper bound. Overall, we have observed that
\begin{align}
    p_\text{crit}(N) + 1 \leqslant m + n < \ell
\end{align}
seems sufficient to guarantee global stability for all compactly supported initial data. 
\\

In fact, we see a feasible route to prove a large-data global stability statement, if one restricts to energy-critical $\ell =  P_\text{crit}(N) + 1$ self-interactions, such that rigorous large-data global stability results apply to the decoupled theory, see~\cref{sec:math:large-data} and, e.g.,~\cite{KenigMerle2006wave}. Our condition would then imply $p_\text{crit}(N) + 1 \leqslant m + n < \ell \leqslant P_\text{crit}(N) + 1$, i.e., that the ghostly interaction can potentially be treated as a subcritical perturbation to an energy-critical global stability proof of the decoupled theory. 

An essential step in large-data global-stability proofs is the absence of nontrivial finite-energy stationary solutions to the respective elliptic equations (obtained by setting time derivatives in~\cref{eq:eoms} to zero). 
This is reminiscent of analogous stationary-state obstructions in non-ghostly potentials.
It may well be that a rigorous proof requires a choice of potential for which one can establish the absence of any nontrivial finite-energy stationary solutions.

\section{Discussion}
\label{sec:discussion}

We have reviewed mathematical theorems, obtained numerical evidence, and presented physical reasoning in favour of a re-assessment of the global stability of classical field theories with ghosts.
\\

First, we have reviewed the mathematical literature from the viewpoint of a theoretical physicist. In particular, we have clarified that proven small-data global stability theorems apply to a class of classical field theories where two opposite-sign (non-tachyonic) scalar fields interact via a non-derivative potential, if the latter are of sufficiently high degree when expanded around the respective vacuum. 

We highlight the significance of this rigorous mathematical fact. The presence of opposite-sign kinetic terms has long been used to rule out any such theory as a candidate for a fundamental theory of nature, even at the classical level. We believe that this prejudice should be dropped in favour of unbiased further scientific assessment. In fact, we are not demanding stronger (or even the same) rigorous stability criteria from our most successful descriptions of nature: For instance, black holes in General Relativity are widely expected to exhibit small-data global stability but, for Kerr black holes, the respective proof still remains to be closed, see~\cref{sec:small-data:derivatives} for references to ongoing work. Nevertheless, most physicists happily accept black holes in General Relativity as a our best description of various astrophysical phenomena\footnote{
A more significant objection against ghostly theories is that no such model has successfully been used to describe significant physical phenomena. On the one hand, this might, of course, be a result of the above (precipitous) prejudice but, on the other hand, any actual physical application remains outstanding.
}.
\\

Second, we have gone beyond small-data and presented a systematic numerical study of large-data spherically-symmetric scattering solutions. While a mathematical proof remains outstanding, we have presented numerical evidence and physical reasoning in favour of a respective large-data global stability conjecture. We will seek collaboration with mathematicians to establish rigorous proof in future work.

The results of our numerical study and the resulting expectations are in line with previous physical insights developed in the context of point-particle systems~\cite{Deffayet:2023wdg} and can be summarised colloquially as follows: In classical field theory, the Ostrogradski instability of ghostly non-derivative interactions between opposite-sign kinetic terms can apparently be quenched by sufficiently dominant self-interactions.

Comparing once more to General Relativity, a similar large-data global stability statement, even restricted to the exterior domain outside any potential apparent horizons, would involve a complete proof of weak cosmic censorship, see, e.g.,~\cite{Wald:1997wa} for review. This is a tall order and has not yet been established. Of course, General Relativity is a much more complex theory than the simple scalar field theories at hand, in particular, with regards to the presence of derivative interactions (see discussion below).
\\

While the current work may significantly alter expectations, several important extensions in the pursuit of further re-assessment of the fundamental viability of higher-derivative theories remain to be explored. In particular, we identify too important directions for future work.

First, we emphasise that all statements in the current paper hold only in absence of derivative interactions. Herein, we refer to a statement at the level of second-order equations of motion. In fact, as discussed in~\cref{sec:small-data:derivatives}, our statements should hold for a class of higher-derivative scalar field theories for which the Lagrangian can be written as a function of the fields and the wave operator. This is because the respective field equations can be trivially order-reduced to a set of nonlinear wave equations with non-derivative interactions. Nevertheless, the impact of other derivative interactions remains to be scrutinised.

Second, it remains unclear whether or not such theories, even if classically stable, can be quantised.
In light of existing claims about catastrophic decay obtained from tree-level Feynman diagrams~\cite{Carroll:2003st,Cline:2003gs}, we add a brief discussion of our personal expectation. In general, one would expect that the tree-level diagrams in a perturbative expansion should correspond to the classical limit. This leads to an apparent tension between the presented global stability results in the nonlinear initial value formulation of the classical field theory and the above statements obtained from tree-level perturbative diagrams. One may conclude that either, the perturbative expansion breaks down, or the quantum theory does not exhibit a classical limit. However, we highlight that the Feynman diagrams in~\cite{Carroll:2003st,Cline:2003gs} correspond to a perturbative expansion of a boundary-value problem in terms of asymptotic states at early and late time. In contrast, classical time evolution is an initial-value problem: A proper assessment of quantum stability should thus rather be extracted from the respective initial-value formalism in perturbative quantum field theory, i.e., from the Schwinger-Keldysh formalism, see, e.g.,~\cite{kamenev2023field} and references therein.
We thus conclude that a proper understanding of the classical limit and its leading quantum corrections remains outstanding and will pursue this line of research in future work.
\\

\paragraph*{Acknowledgements.}
The author wishes to thank C.~Deffayet, S.~Mukohyama, and A.~Vikman for discussions on related work in the context of ghostly field theories; P.~Figueras and A.~Kovacs for discussions on related work in the context of effective field theory; and Ramiro Cayuso for discussion about the implementation of regularity conditions in radial evolution codes. 
The author also thanks C.~Deffayet, F.~Merle, J.~Szeftel, M.~Taylor, and T.~Wiseman for discussions on small-data and large-data global stability, as well as T.~Colas, E.~Pajer, C.~de~Rham and A.~Tolley for discussions about the Schwinger-Keldysh formalism. 
All numerical results have been obtained with up to $2^{15}$ spatial grid points and have been performed on a single node (16 cores at 2.8 GHz each and up to 1 TB of RAM) of the Thanos cluster at LPENS.


\appendix

\section{Growing modes and diagonalisation of higher-derivative systems}
\label{app:growing_modes}

In this appendix, we discuss how to diagonalise the order-reduced 2nd-order equations of motion for the linear part of a higher-derivative system. If there are additional nonlinearities, these will generically not decouple. We also use this opportunity to explicitly show that exponentially growing modes of the linear problem only arise in the case of tachyonic frequencies / mass terms. Attributing this exponential growth to the higher-derivative (ghostly) nature of the equations is misleading since non-tachyonic parameter choices do not exhibit any exponentially growing modes.

\subsection*{Point-particle motion}

As a first example, we consider the higher-derivative point-particle example discussed in~\cite{Woodard:2015zca} in which a harmonic oscillator of frequency $\omega$ is supplemented by a higher-derivative correction with respective parameter $\epsilon$. The respective Lagrangian reads
\begin{align}
L &= -\frac{\epsilon m}{2 \omega^2} \ddot{x}^2 + \frac{m}2 \dot{x}^2
- \frac{m\omega^2}2 x^2 \; , 
\label{HDO}
\end{align}
and the equations of motion are
\begin{align}
0 &= -m \Bigl[ \frac{\epsilon}{\omega^2} \ddddot{x} + \ddot{x} + 
\omega^2 x\Bigr] \; . \label{HDE}
\end{align}
The example is trivial in a sense that it can be decoupled and there are no genuine interactions. Still, it shows that exponential growth arises due to tachyonic frequencies, not due to the higher-derivative nature. 

Setting, for simplicity $m=1$, the system has a general (non-degenerate) solution
\begin{align}
    x(t) &=  
    \sum_{i=1}^4 C_i \exp(\lambda_i t) 
    \; ,
    \label{gensol}
\end{align}
where the frequencies $\lambda_i$ are determined as the roots of the characteristic polynomial $\epsilon/\omega^2\,\lambda^4 + \lambda^2 + \omega^2\lambda = 0$, i.e.,
\begin{align}
    \lambda_{i} \equiv \pm \sqrt{
        \frac{
            \left(1 \pm \sqrt{1-4 \epsilon}\right)\omega^2
        }{
            2 \epsilon
        } 
    }
    \;,
\end{align}
and $i=1,2,3,4$ labels the four possible sign combinations. 
There are exponentially growing modes if the respective roots have positive real parts. However, this is not a consequence of the higher-derivative nature of the equations but rather of tachyonic frequencies.
In fact, for $0<\epsilon<1/4$ and $\omega\in\mathbb{R}$, the equations still contain higher derivatives but all roots have vanishing real part and the equations thus describe two decoupled non-tachyonic oscillators.

We may also rewrite the equations in second-order form
\begin{align}
    \ddot{x} &\equiv y\;,
    \\*
    \ddot{y}&= - \frac{\omega^2}{\epsilon}y - \frac{\omega^4}{\epsilon}x\;,
\end{align}
which in turn can be diagonalised (assuming non-degeneracy of the eigenvalues) to
\begin{align}
    \ddot{u} &= \omega_+^2\,u\;,
    \\
    \ddot{v}&= \omega_-^2\,v\;,
\end{align}
with $x=u+v$ and $y=\omega_+^2\,u + \omega_-^2\,v$ and
$
\omega_{\pm}^2 = \frac{
            \left(1 \pm \sqrt{1-4 \epsilon}\right)\omega^2
        }{
            2 \epsilon
        }
$. The decoupled system again shows that exponential vs. oscillatory behaviour is determined by the respective frequencies being tachyonic or not.
Finally, note that if (some of) the roots are degenerate, one can get secular polynomial growth, e.g., for $\ddddot{x}=0$, the general solution is obviously $x(t) = \sum_{i=1}^4\,C_i t^{i-1}$.

\subsection*{Higher-derivative field theory}

The same algebraic relations also hold for the corresponding field theory with a Lagrangian
\begin{align}
L &= -\frac{\epsilon}{2 M^2} \phi\,\Box^2\phi + \frac{1}{2} \phi\,\Box\,\phi
- \frac{M^2}{2} \phi^2 \; , 
\label{HDO}
\end{align}
and field equations
\begin{align}
0 &= \frac{\epsilon}{M^2} \Box^2\phi + \Box\phi + 
M^2\,\phi \; , \label{HDE}
\end{align}
or in decoupled (non-degenerate) second-order form
\begin{align}
    \Box\,\phi_+ &= M_+^2\,\phi_+\;,
    \\
    \Box\,\phi_- &= M_-^2\,\phi_-\;,
\end{align}
where, again, $\phi=\phi_+ + \phi_-$ and $\psi=M_+^2\,\phi_+ + M_-^2\,\phi_-$ and
$
M_{\pm}^2 = 
\left(1 \pm \sqrt{1-4 \epsilon}\right)M^2
/\left(2 \epsilon\right)
$.
Once more, we see that tachyonic (non-tachyonic) masses imply the presence (absence) of exponentially growing modes and, for $M\in\mathbb{R}$ and $0<\epsilon<1/4$, there are no exponentially growing modes. In fact, for $M\in\mathbb{R}$ and $0<\epsilon<1/4$, the system is stable.

In the degenerate case $\Box^2\phi=0$, one finds again secular---in this case, linear---growth. For instance, this can be seen by rewriting the degenerate case in second-order form as
\begin{align}
    \Box\,\phi &\equiv \psi\;,
    \\    
    \Box\,\psi &= 0\;.
    \label{eq:field-theory_decoupled}
\end{align}
The general solution can be obtained (using Fourier-space methods) by, first, solving the second equation which is nothing but a homogeneous wave equation and, second, solving the first equation on that previous solution via homogoneous and particular part. The result is
\begin{align}
\phi(t, \mathbf{x}) =\; & 
\underbrace{
\int d^3k\, \left[ c(\mathbf{k})\, e^{-i\left(|\mathbf{k}| t - \mathbf{k} \cdot \mathbf{x}\right)} + d(\mathbf{k})\, e^{+i\left(|\mathbf{k}| t - \mathbf{k} \cdot \mathbf{x}\right)} \right]
}_{\text{homogeneous solution to } \Box \phi = 0} \nonumber \\
&+ 
\underbrace{
\int d^3k\, \left[ \frac{i a(\mathbf{k})}{2|\mathbf{k}|}\, t\, e^{-i\left(|\mathbf{k}| t - \mathbf{k} \cdot \mathbf{x}\right)} 
+ \frac{-i b(\mathbf{k})}{2|\mathbf{k}|}\, t\, e^{+i\left(|\mathbf{k}| t - \mathbf{k} \cdot \mathbf{x}\right)} \right]
}_{\text{particular solution to } \Box \phi = \psi}
\;.
\end{align}
This shows the secular growth as well as the absence of exponential growth.



\bibliography{References}
    
\end{document}